\documentclass[12pt]{article}

\usepackage[utf8]{inputenc}
\usepackage{geometry}
\usepackage{graphicx}
\usepackage{array}
\usepackage{slashed}
\usepackage{amsmath}
\usepackage{amsfonts}
\usepackage{amssymb}
\usepackage{physics}
\usepackage{mathtools}
\usepackage{placeins}
\usepackage[usenames, dvipsnames]{color}
\usepackage{caption}
\usepackage{subcaption}
\usepackage{dsfont}
\usepackage[cm]{fullpage}
\usepackage{cite}
\usepackage{epstopdf}
\usepackage{comment}
\usepackage{hyperref}
\usepackage{enumitem}
\usepackage{cancel}
\usepackage{mathrsfs}

\usepackage{cleveref}
\crefname{section}{§}{§§}
\Crefname{section}{§}{§§}
\usepackage{booktabs} 

\numberwithin{equation}{section}

\def\p{\partial}

\def\0{{(0)}}
\def\1{{(1)}}
\def\2{{(2)}}

\def\<{\langle }
\def\>{\rangle }

\newcommand{\bea}{\begin{eqnarray}}

\newcommand{\eea}{\end{eqnarray}}

\newcommand{\be}{\begin{equation}}
\newcommand{\ee}{\end{equation}}
\newcommand{\ba}{\begin{align}}
\newcommand{\ea}{\end{align}}

   \makeatletter
  \let\over=\@@over \let\overwithdelims=\@@overwithdelims
  \let\atop=\@@atop \let\atopwithdelims=\@@atopwithdelims
  \let\above=\@@above \let\abovewithdelims=\@@abovewithdelims
\renewcommand\section{\@startsection {section}{1}{\z@}%
                                   {-3.5ex \@plus -1ex \@minus -.2ex}
                                   {2.3ex \@plus.2ex}%
                                   {\normalfont\large\bfseries}}

\renewcommand\subsection{\@startsection{subsection}{2}{\z@}%
                                     {-3.25ex\@plus -1ex \@minus -.2ex}%
                                     {1.5ex \@plus .2ex}%
                                     {\normalfont\bfseries}}

\newcommand{\beq}{\begin{equation}}
\newcommand{\eeq}{\end{equation}}
\newcommand{\beqa}{\begin{eqnarray}}
\newcommand{\eeqa}{\end{eqnarray}}
\newcommand{\beqar}{\begin{eqnarray*}}

\def\[{\[}
\def\]{\]}

\newcommand{\bd}[1]{\begin{fmffile}{#1}\begin{fmfgraph*}}
\newcommand{\ed}{\end{fmfgraph*}\end{fmffile}}

\begin{document}

\begin{titlepage}

\unitlength = 1mm~\\
\vskip 1cm
\begin{center}

{\LARGE{\textsc{From Free-Fermionic Constructions to Orbifolds and back}}}

\vspace{0.8cm}
Ioannis Florakis\,{}\footnote{{\tt iflorakis@uoi.gr}} and John Rizos\,{}\footnote{{\tt irizos@uoi.gr}}

\vspace{1cm}

{\it  Department of Physics, University of Ioannina\\ GR45110 Ioannina, Greece 

}

\vspace{0.8cm}

\begin{abstract}
We systematically develop the explicit map between string vacua constructed in the Free Fermionic Formulation and their $\mathbb Z_2^N$ toroidal orbifold counterparts.
We illustrate the map in various example classes of models, including cases relevant for string phenomenology, as well as in theories where space-time supersymmetry is broken by the stringy Scherk-Schwarz mechanism.

\end{abstract}

\setcounter{footnote}{0}

\vspace{1.0cm}

\end{center}

\end{titlepage}

\pagestyle{empty}
\pagestyle{plain}

\def\vx{{\vec x}}
\def\p{\partial}
\def\po{$\cal P_O$}

\pagenumbering{arabic}

\tableofcontents



\section{Introduction}

Since its appearance in the 80's, the free-fermionic formulation \cite{Antoniadis:1985az,Kawai:1986ah,Kawai:1986va,Antoniadis:1986rn,Antoniadis:1987wp} has provided a powerful framework for constructing string vacua, particularly useful for model building. In free-fermionic constructions one exploits two-dimensional bosonization to consistently replace the CFT of compact worldsheet scalars at special points in moduli space, by a CFT of auxiliary free fermions. Aside from the non-compact bosonic coordinates associated to four-dimensional Minkowski spacetime, the remaining worldsheet d.o.f. consist entirely of free fermions which can realise worldsheet supersymmetry non-linearly among themselves \cite{Antoniadis:1985az}. Higher loop modular invariance and factorisation then result in a set of constraints that can be solved \cite{Antoniadis:1986rn,Antoniadis:1987wp} in terms of relatively simple construction rules to give rise to consistent vacua. These are defined in terms of their worldsheet fermion content -- namely, by  choices of boundary conditions (spin structures) for the free fermions, as well as by a set of Generalised GSO (GGSO) projections. The upshot is that the algorithmic nature of the construction's rules is particularly suited for computer-aided scans, and has given rise to several semi-realistic models that became the subject of considerable study in the literature (for instance, see \cite{Antoniadis:1989zy,Faraggi:1989ka,Antoniadis:1990hb,Leontaris:1999ce,Faraggi:2004rq,Assel:2010wj,Faraggi:2014hqa,Faraggi:2017cnh,Abel:2014xta,Faraggi:2021mws}).

Despite its advantages for model scanning, the free-fermionic formalism is not directly suited for dealing with questions related to the dependence of masses (or other quantities of interest) on the compactification moduli. Indeed, the ``fermionization" procedure requires compact bosons to sit at special loci in moduli space, so that moduli describing the size and shape of the compactification space are set to special values, typically of order one in string units. This is often sufficient for phenomenological purposes, assuming the vacuum preserves at least some target-space supersymmetry. In cases when supersymmetry is broken, however, the nature of the breaking can be obscured within a purely free-fermionic formalism. In such cases, the moduli can be reinstated by introducing Thirring interactions \cite{Bagger:1986cd,Chang:1988xs,Chang:1988hf}, or by mapping the theory to toroidal orbifold constructions \cite{Dixon:1985jw,Dixon:1986jc,Ibanez:1987pj} and deforming it there. In this work we shall focus on the latter.

Breaking supersymmetry in heterotic free-fermionic constructions amounts to picking the boundary conditions of the free fermions and the corresponding GGSO coefficients such that the theory preserves strictly $\mathcal N=1$ worldsheet supersymmetry, while breaking the would-be $\mathcal N=2$ worldsheet SCFT enhancement. In orbifold compactifications, the breaking of supersymmetry may be considered either spontaneous or explicit, depending on whether the orbifold responsible for the breaking acts  freely or not. In the former case, there are no fixed points and the corresponding gravitini are not projected out of the spectrum but rather become massive. In the latter case, there are fixed points, the corresponding gravitini are projected out of the spectrum, and the low energy theory has no memory of the original supersymmetry. However, although the interplay between $\mathbb Z_2$ rotations (twists) and translations (shifts) is transparent in bosonic coordinates, a free action can be sometimes obscured in the free-fermionic setup, where it enters also into the choice of GGSO coefficients.

To answer such questions, it becomes important and necessary to deform the theory away from the fermionic point and bring the compactification moduli $\{\mu\}$ to generic values, hence, allowing \emph{e.g.} a study of the gravitino mass scale, $m_{3/2}$, as a function of the $\mu$'s. Indeed, by re-bosonizing the auxiliary fermions, it is often desirable to re-interpret the theory as a toroidal $\mathbb Z_2^N$ orbifold. In the latter representation, compactification moduli enter the one-loop partition function through Narain lattice sums, possibly amended to account for orbifold twists and shifts. VEVs of the compactification moduli then explicitly enter the tree-level masses of states in the theory, so that analysing their effects becomes, in principle, straightforward.

Assuming that the choice of boundary conditions of the free fermions is compatible with spacetime supersymmetry, these considerations lead to a natural procedure for classifying the GGSO coefficients of non-supersymmetric free fermionic theories, depending on whether the breaking they induce is spontaneous, as in Scherk-Schwarz compactifications \cite{Scherk:1978ta,Scherk:1979zr}, or explicit. Starting with a free-fermion theory and, assuming that the explicit map to an orbifold representation is known, the theory can be deformed away from the fermionic point and the moduli dependence of the states can be extracted. In particular, one may check whether $m_{3/2}(\mu)\to 0$ vanishes,  as we approach the boundary of the perturbative moduli space. This provides a selection criterion --  it identifies choices of GGSO projections that are compatible with spontaneous supersymmetry breaking. These cases correspond to the stringy generalisation of the Scherk-Schwarz mechanism \cite{Rohm:1983aq,Kounnas:1988ye,Ferrara:1988jx}, inducing supersymmetry breaking at a scale $m_{3/2}\sim 1/R$, inversely proportional to the size $R$ of the internal space, while still preserving an exact CFT realisation. Indeed, the stringy Scherk-Schwarz may be formulated either as a coordinate-dependent compactification \cite{Kounnas:1989dk} (\emph{i.e.} a boost of the fermionic and bosonic charge lattices) or, equivalently, as a freely-acting $\mathbb Z_2$ orbifold of the original theory \cite{Kiritsis:1996xd,Kiritsis:1997ca,Kiritsis:1998en}. 

The idea that a large class of free-fermionic models, including some of the most relevant ones for string phenomenology, corresponds to toroidal $\mathbb Z_2^N$ orbifolds is certainly not new. The connection has been indeed discussed in the literature in several works \cite{Chamseddine:1989mz, Faraggi:1993pr, Kiritsis:1997ca, Berglund:1998eq, Berglund:1998rq, Faraggi:2002qh, Faraggi:2004rq, Donagi:2004ht, Faraggi:2006bs, Donagi:2008xy, Faraggi:2011aw, florakis:tel-00607408, Athanasopoulos:2016aws, Florakis:2016ani, Florakis:2021bws, Florakis:2022avh, Avalos:2023mti, Avalos:2023ldc}. At the very heart of the correspondence lies, of course, two-dimensional bosonization, along with the fact that order-two rotations of bosonic worldsheet coordinates have a simple realisation in terms of their free fermionic counterparts. Together with the fact that a square lattice is invariant under $\mathbb Z_2$ boundary conditions, it becomes more or less apparent that the construction of a large class of vacua in the free-fermionic framework must admit an equivalent orbifold representation. Although the correspondence between the two formalisms is, in principle, clear, the task of exactly matching the two formulations can easily become a daunting technical task, especially when semi-realistic models are concerned.

Indeed, despite early efforts in \cite{florakis:tel-00607408} and the subsequent investigation in \cite{Athanasopoulos:2016aws}, a general systematic map between the two formulations is still not complete. The present work aims to fill some of these gaps, with particular interest in incorporating the stringy Scherk-Schwarz realisations. The main difficulty in establishing a precise map is two-fold. The first is related to the different ways the free-fermionic and orbifold constructions package together the boundary conditions of the various worldsheet degrees of freedom. It can be resolved by careful book-keeping and bosonization, together with a comparison of the corresponding one-loop partition functions, after accounting for additional phases arising from the different conventions and periodicities of the theta functions induced by the difference in packaging. Sorting this out results in a joint form of the partition function of the two constructions, that is expressed entirely in terms of theta functions, whose characteristics match term by term. Indeed, when expressed solely in terms of theta functions, the modularity properties of the partition function are unchanged if one consistently includes phases coupling together boundary conditions of different fermions, provided that they are themselves invariant under the representation of modular transformations on the space of spin structures. The freedom to introduce such ``modular invariant" phases in an orbifold-like partition function can be seen to be in one-to-one correspondence with the freedom to choose the GGSO phases in the free fermionic formulation. The second difficulty one needs to resolve is related to the matching of the GGSO phases with the ``modular invariant" cocycles of the orbifold-like partition function, and then consistently replacing the theta functions corresponding to the fermionization of the internal space coordinates with (twisted/shifted) Narain lattice sums while keeping track of the ``transmutation" of fermionic boundary condition labels into freely-acting orbifold ones. This last step may lead to different equivalent representations of the orbifold theory.

The paper is organised as follows: in Section \ref{FFconstruction} we briefly review the free fermionic construction rules and set much of the notation that will be used throughout. In Section \ref{MExpample}, we present the salient features of the map in the simplest toy model involving only two basis vectors. This serves as a suggestive introduction to the machinery that will be employed in later sections. In Section \ref{GenericMap}, we present the general procedure of the map and in Section \ref{SymExample} we proceed to apply it to a class of symmetric $T^6/\mathbb{Z}_2\times\mathbb{Z}_2$ orbifolds, which we treat in considerable detail. We conclude in Section \ref{LatticeReps} with a discussion of the redundancy present in certain choices of phases and provide examples of how it can be lifted by a careful examination of the lattices.


\section{The free fermionic construction rules}
\label{FFconstruction}

We focus on heterotic vacua in four-dimensional Minkowski spacetime, which is the most relevant case for string phenomenology. The left-moving (RNS) sector degrees of freedom realise an $\mathcal N=1$ SCFT, which will necessarily include $4$ spacetime bosons $X_L^\mu$ and fermions $\psi^\mu$, the reparametrisation ghosts $b,c$, and the superghosts $\beta,\gamma$.  For the purposes of our analysis, the net effect of the $b,c$ and $\beta,\gamma$ systems is essentially the cancellation of the longitudinal components of $X^\mu$ and $\psi^\mu$. In what follows, we will effectively work in the light-cone and restrict $\mu$ to run only over the transverse spacetime directions. The total left-moving central charge may be cancelled by the introduction of 18 real fermions, which we denote $\chi^I, y^I,\omega^I$ for $I=1,2,\ldots 6$. In this setup, worldsheet supersymmetry must be realised non-linearly as explained in \cite{Antoniadis:1985az}, eventually requiring that the 18 fermions transform in the adjoint representation of a compact, semi-simple Lie group. Here we consider the simplest case exploited in free-fermionic model building corresponding to the choice $\text{SU}(2)^6$, where the worldsheet supercurrent takes the form
\begin{equation}
	T_F(z) = i\psi^\mu \partial X_\mu + i\sum_{I=1}^6 \chi^I y^I \omega^I \,.
	\label{supercurrent}
\end{equation}
The right-moving worldsheet is bosonic and includes the transverse spacetime bosons $X_R^\mu$, the right-moving ghosts $\tilde b,\tilde c$ and 44 real fermions, as required by the cancellation of the total right-moving central charge.

Even before discussing boundary conditions, the counting of the fermionic worldsheet fields can be explained by comparing, for instance, with the ten-dimensional heterotic strings toroidally compactified on $T^6$. The six-dimensional compact space is spanned by the internal worldsheet coordinates $X^I$ which, at the fermionic radius $R=\sqrt{\alpha'/2}$, can be fermionized as $\partial X^I(z)/\sqrt{\alpha'} \simeq y^I \omega^I$  (we suppress the normal ordering symbol). This accounts for 12 real left-moving  fermions, while the remaining 6 are simply the original RNS superpartners $\chi^I(z)$ of the worldsheet bosons $X^I_L$. The right-moving compact coordinates $X^I_R$ are similarly fermionized in terms of 12 real fermions $\bar y(\bar z)$ and $\bar\omega(\bar z)$, while the remaining 32 real right-moving fermions $\bar\lambda^a(\bar z)$ are identified as the auxiliary fermions entering the usual fermionic formulation of the heterotic string. The set of $\lambda^a$'s can be complexified in pairs and, adopting a notation commonly used in much of the model-building literature, they can be labelled $\bar\psi^1,\bar\psi^2,\ldots,\bar\psi^5$, $\bar\eta^1, \bar\eta^2, \bar\eta^3$, and $\bar\phi^1,\ldots,\bar\phi^8$. This notation is particularly useful for models containing an $\text{SO}(10)$ gauge group factor, though at this stage the theory is rather general.

Focusing on rational CFTs, the boundary condition of each worldsheet fermion $f$, parallel transported around a non-contractible loop $\alpha$ in a given worldsheet topology, is described by a rational number $\alpha(f)\in (-1,1]$,
\begin{equation}
	f\to -e^{i\pi\alpha(f)} \,f \,,
\end{equation}
with the special case $\alpha(f)=0,1$ corresponding to anti-periodic (NS) and periodic (R) fermions, respectively. We may then identify $\alpha$ with a vector of Lorentzian signature $(20,44)$ whose elements are the boundary condition parameters $\alpha(f)$ for all worldsheet fermions in the theory. For example, using a canonical order often encountered in the literature,
\begin{equation}
	\begin{split}
		\alpha =& [\, \alpha(\psi^\mu), \alpha(\chi^{1}), \ldots, \alpha(\chi^{6}), \alpha(y^{1}), \ldots, \alpha(y^6), \alpha(\omega^1),\ldots,\alpha(\omega^6) ; \\
			    &	\alpha(\bar y^1),\ldots,\alpha(\bar y^6), \alpha(\bar\omega^1),\ldots, \alpha(\bar\omega^6), \alpha(\bar\psi^1),\ldots,\alpha(\bar\psi^5), \\
			    &\alpha(\bar\eta^1),\ldots,\alpha(\bar\eta^3), \alpha(\bar\phi^1),\ldots, \alpha(\bar\phi^8) \,]\,,
	\end{split}
	\label{DefAlpha}
\end{equation}
where the semicolon at the end of the first line separates the left and right movers\footnote{Depending on the boundary condition assignments, it is often advantageous to complexify fermion pairs. In such cases, the vector $\alpha$ will contain the boundary conditions shared by the complexified pair.}. In cases when only periodic or anti-periodic boundary conditions appear, it is sometimes customary to represent the vector $\alpha$ directly as the set of those fermions that are periodic.

When the worldsheet has the topology of a torus, there are two non-contractible cycles $\alpha$ and $\beta$, so that a choice of spin structure corresponds to a pair of boundary condition vectors $(\alpha,\beta)$.
The one-loop vacuum amplitude of the theory is expressed as a sum over such spin structures belonging to a set $\Xi$ 
\begin{equation}
	\mathscr{Z} = \int_{\mathcal F} \frac{d^2\tau}{\tau_2^3}\,\frac{1}{\eta^{12}\,\bar\eta^{24}}\,\frac{1}{|\Xi|} \sum_{\alpha,\beta\in\Xi} C\left[\alpha \atop \beta\right]\,\prod_{f-\text{real}} \Theta^{1/2}\left[\alpha(f)\atop\beta(f)\right](0;\tau) \prod_{f- \text{complex}} \Theta\left[\alpha(f)\atop\beta(f)\right](0;\tau)  \times \text{(\ldots)} \,,
	\label{FFpartition}
\end{equation}
where $\tau$ is the complex structure of the worldsheet torus, $\mathcal F$ is the fundamental domain, and $C\left[\alpha\atop\beta\right]$ are the $\tau$-independent GGSO coefficients, that will be identified as pure phases in what follows. In the above expression, 
$|\Xi|$ denotes the cardinality of $\Xi$, while the products run over the real and complex worldsheet fermions, respectively. Furthermore $\eta(\tau)$ is the Dedekind eta function
\begin{equation}
	\eta(\tau) = q^{1/24}\prod_{n>0}(1-q^n)\,,
\end{equation}
with $q=\exp(2\pi i\tau)$ being the nome, while $\Theta$ is the Jacobi theta function with characteristics in the Mumford convention
\begin{equation}
	\Theta\left[\phi \atop \chi\right](z;\tau) = e^{-i\pi \phi\chi/2} \sum_{n\in\mathbb Z} q^{\frac{1}{2}(n-\phi/2)^2} e^{2\pi i (z-\chi/2)(n-\phi/2)} \,.
	\label{MumfordTheta}
\end{equation}
The right-moving counterpart, involving products of the right-moving theta factors, is completely analogous and is denoted by the ellipses in \eqref{FFpartition}. 
As shown in \cite{Antoniadis:1986rn,Antoniadis:1987wp}, multi-loop modular invariance implies that that the only non-vanishing spin structure contributions correspond to pairs of elements of an (abelian) additive group 
\begin{equation}
	\Xi = \left\{\alpha \Big| C\left[\alpha\atop 0\right] =(-1)^{\alpha(\psi^\mu)} \right\}\,,
\end{equation}
which must necessarily contain the vector $\mathds{1}$, corresponding to periodic boundary conditions for all fermions.
Furthermore, the GGSO coefficients must satisfy the conditions
\begin{equation}
	\begin{split}
		C\left[\alpha \atop \beta\right] &= e^{\frac{i\pi}{4}(\alpha\cdot\alpha+\mathds{1}\cdot\mathds{1})} \,C\left[\alpha \atop \beta-\alpha+\mathds{1}\right] \,,\\
		C\left[\alpha \atop \beta\right] &= e^{\frac{i\pi}{2}\alpha\cdot\beta}\, C\left[\beta \atop -\alpha \right] \,, \\
		C\left[\alpha \atop \beta+\gamma \right] & = (-1)^{\alpha(\psi^\mu)}\,C\left[\alpha \atop \beta\right]\,C\left[\alpha \atop \gamma\right] \,.
	\end{split}
	\label{modularity}
\end{equation}
The dot product has a Lorentzian signature, with left-movers contributing with plus signs and right-movers with minus signs, while terms corresponding to real fermions are weighted with an additional factor of $1/2$ compared to complex ones. 
In the rational case of interest here, the group $\Xi$ is finitely generated, \emph{i.e.} $\Xi = \mathbb{Z}_{N_1}\oplus \ldots \oplus \mathbb{Z}_{N_k}$, and the GGSO coefficients are pure phases. This implies that there exists a basis $\mathscr{B}=\{\beta_i\}_{i=1}^k$ generating $\Xi$, such that the vectors $\beta_i$ are linearly independent in the sense that $\sum_i m_i \beta_i=0$ if and only if $m_i=0\,(\text{mod}\,N_i)$ for all $i=1,\ldots,k$. It is convenient to pick $N_1=2$ so that $\beta_1=\mathds{1}$. Following \cite{Antoniadis:1987wp}, the basis vectors are then required to satisfy 
\begin{equation}
	N_i\,\beta_i\cdot\beta_i = 0\, (\text{mod}\, 8)\ ,\quad N_{ij}\,\beta_i\cdot\beta_j = 0\, (\text{mod}\, 4)\,,
	\label{FFrules}
\end{equation}
where $N_{ij}= \text{l.c.m.}(N_i,N_j)$, along with the condition that the number of real fermions which are simultaneously periodic in any combination of four basis vectors is even. Furthermore, the boundary condition vectors must preserve the worldsheet supercurrent \eqref{supercurrent}, implying the additional constraint
\begin{equation}
	\beta_i(\psi^\mu)+\beta_i(\chi^I) + \beta_i(y^I)+ \beta_i(\omega^I) = 0 \,(\text{mod}\,2) \,,
\end{equation}
for all $i=1,\ldots,k$ and $I=1,\ldots,6$. Since vectors $\xi\in\Xi$ can be expanded as $\xi = \sum_i m_i \beta_i$ with $m_i\in\mathbb{Z}_{N_i}$, the modularity conditions \eqref{modularity} may be utilised to express any GGSO coefficient in terms of the GGSO matrix coefficients $C_{ij} \equiv C\left[ \beta_i \atop \beta_j \right]$ corresponding to the elements of the basis $\mathscr{B}$. The latter $k^2$ elements are not all independent, since they too are constrained to satisfy the constraints \eqref{modularity}. For purely real boundary conditions ($N_i=2$ for all $i$), they may be expressed in terms of $k(k-1)/2+1$ independent phases, which may be chosen to be $C_{11}$ and $C_{ij}$ with $i<j$. 

Each choice of GGSO coefficients satisfying these requirements corresponds to a consistent string theory, with the Hilbert space being given by
\begin{equation}
	\mathscr{H}= \bigoplus_{\alpha\in\Xi} \left\{ \prod_{i=1}^k  P_{\alpha,i} \right\} \mathscr{H}_\alpha\,.
\end{equation}
Here, $\mathscr H_\alpha$ is the Hilbert space quantised with the boundary conditions $\alpha$, while $P_{\alpha, i}$ is the GGSO operator that projects onto states satisfying 
\begin{equation}
	e^{i\pi\beta_i \cdot F_\alpha}=(-1)^{\alpha(\psi^\mu)}C\left[\alpha\atop\beta_i\right]^*\,,
	\label{FFprojector}
\end{equation}
with $F_\alpha$ being the worldsheet fermion number (it counts fermions $f$ with a plus sign, their conjugates $f^*$ with a minus sign, while acting as a chirality operator on the R vacuum).


\section{Preparatory observations and a minimal example}
\label{MExpample}

In this work we discuss the map between the four-dimensional heterotic free fermionic constructions of Section \ref{FFconstruction} and compactifications on toroidal orbifolds. Although the additive group $\Xi$ was taken to be a direct sum of $\mathbb Z_{N_i}$ factors, for concreteness we henceforth take all $N_i=2$ so that all fermions are either periodic or anti-periodic, \emph{i.e.} $\alpha(f)=0,1$. This effectively corresponds to orbifold groups composed of at most $\mathbb Z_2$ factors. In view of the bosonisation relation $y^I \omega^I \simeq \partial X^I/\sqrt{\alpha'}$, it is clear that the boundary condition assignments entering the basis $\mathscr{B}=\{\beta_i\}$ correspond to different actions on the internal space coordinates $X^I$. For example, a $\mathbb Z_2$ rotation $X^I\to-X^I$ may be realised in the space of auxiliary fermions as $y^I\to-y^I$, $\omega^I\to +\omega^I$. In those cases, we will conventionally choose to transform only the $y$'s while leaving $\omega$'s inert. Similarly, the order-2 translation, $X^I \to X^I +\pi R^I$, is realised at the fermionic point, $R^I=\sqrt{\alpha'/2}$, as the simultaneous action $y^I\to -y^I$, $\omega^I\to -\omega^I$. 

Under translations, the left-moving currents $\partial X^I$ are left invariant and, hence, by worldsheet supersymmetry, so should the RNS fermions $\chi^I$. Spacetime supersymmetry is unaffected by a pure translation, as one may easily check \emph{e.g.} by inspection of the vertex operators of the gravitini. A rotation, instead, acts non-trivially on the $\chi^I$'s and, hence, also on the spin fields entering the gravitini vertex operators. Modding out by $\mathbb Z_2$ rotations in orbifold theories typically projects out some of the gravitini and explicitly breaks a number of supersymmetries (a single $\mathbb Z_2$ rotation on a $T^4$ breaks half of them). 

In orbifold theories, the worldsheet degrees of freedom initially have certain boundary condition assignments, preserving some discrete symmetries. Modding out by the latter results in a truncation of the original Hilbert space down to the subspace invariant under the orbifold action, while simultaneously extending it by the addition of twisted sectors. From the path integral point of view, the one-loop partition function can be obtained by calculating the vacuum amplitude of the worldsheet fields with the original boundary conditions along the two non-contractible cycles of the worldsheet torus, while further allowing for twisted boundary conditions in accord with the orbifold action and summing over all twists. 

The same twisting of boundary conditions also arises in free fermionic constructions, as we now explain. First, notice that the presence of at least one massless gravitino in the original theory implies that $\Xi$ must contain the boundary condition vector $S=\{\psi^\mu, \chi^1,\ldots,\chi^6\}$, which we can choose as one of our basis elements. In the $S$ sector, the corresponding Hilbert space $\mathscr H_S$ has a degenerate Ramond vacuum $|S\rangle \sim |\pm\pm\pm\pm\rangle$, transforming as a Weyl spinor of $\text{SO}(8)$, either with an even or an odd number of minus signs. Acting on this with the right-moving $\bar\partial X^\mu$ oscillators then results in spin $3/2$-fields, identified as the 4 gravitini corresponding to the maximal $\mathcal N=4$ supersymmetry of heterotic compactifications on $T^6$ (and their anti-particles). 

Consider then the minimal example of basis $\mathscr{B}=\{\mathds{1}, S\}$, generating the additive group with four elements, $\Xi=\{0,\mathds{1}, S,\mathds{1}+S\}$. For reasons that will become clear shortly, it is convenient to use instead the basis $\mathscr{B}'=\{S,\mathds{1}+S\}$, and expand any element $\xi\in\Xi$ into the linear combination
\begin{equation}
	\xi = a S + m (\mathds{1}+S) ~~ (\text{mod}\,2)\,,
	\label{Expansion1}
\end{equation}
with $a,m=0,1$ being the components of $\xi$ with respect to $\mathscr{B}'$. Observe that $\mathscr{B}'$ is set up such that its basis vectors\footnote{In the case of purely periodic or anti-periodic boundary conditions, where boundary condition vectors can be equivalently described as sets of periodic fermions,  vector addition (modulo 2) corresponds to the symmetric difference of the corresponding sets, \emph{i.e.} $\alpha+\alpha' \,(\text{mod}\,2) \approx (\alpha\backslash \alpha')\cup(\alpha' \backslash\alpha)$.} 
 cover the set of all fermions, $S\cup(\mathds{1}+S)=\mathds{1}$, and are disjoint, $S\cap(\mathds{1}+S)=\varnothing$;  the latter implying that they are also orthogonal with respect to the dot product introduced in Section \ref{FFconstruction}, \emph{i.e.} $S\cdot(\mathds{1}+S)=0$.   We can think of $S$ (more precisely, its conjugate parameter $a$) as describing the common boundary conditions of the RNS fermions, $\psi^\mu$ and $\chi^I$, while $\mathds{1}+S$ (respectively, the parameter $m$) describes the (common) boundary conditions of all remaining fermions. If we now extend $\Xi$ by introducing additional basis elements, we can interpret those new vectors with a non-trivial overlap with $S$ as `twisting' the boundary conditions (of some) of the $\psi^\mu$, $\chi^I$ fermions. The same interpretation clearly applies also to the remaining fermions.

It is useful to classify such extensions depending on the overlap of the additional basis vectors with $S$ since, their effect on spacetime supersymmetry is different.
 Consider first enlarging the basis by a new element $\beta\in\Xi$, such that $\beta\cap S=\varnothing$, \emph{i.e.} associated to some boundary condition `twist' that does not transform the $\psi^\mu$'s nor the $\chi^I$'s. The corresponding projector \eqref{FFprojector} will preserve all 4 gravitini if $C\left[S\atop \beta\right]=-1$; otherwise it projects out all of them. Whether this breaking is explicit or spontaneous is not immediately apparent and we postpone the relevant discussion for later.
 
 Now consider instead a boundary condition vector $\beta'\in\Xi$ that does contain some of the $\chi^I$'s, namely $\beta'\cap S\neq\varnothing$. This can be interpreted as twisting the boundary conditions of (some of) the $\chi^I$'s, since the additive group $\Xi$ will now contain also the sector $S+\beta'$. The expansion of an arbitrary $\xi\in\Xi$ then reads
\begin{equation}
	\xi = aS+m (\mathds{1}+S)+ h \beta' ~~ (\text{mod}\,2)\,,
	\label{Expansion2}
\end{equation}
where $\beta'$ (more precisely, the conjugate parameter $h =0,1$) is exactly identified with the $\mathbb Z_2$ twist. In fact, if $\xi$ corresponds to the ``$\alpha$"-cycle of the worldsheet torus (upper characteristic), the corresponding $h$ is identified with the $\mathbb Z_2$ parameter labelling the untwisted and twisted sectors of the corresponding orbifold, respectively. Similarly, for the ``$\beta$"-cycle (lower characteristic), the analogue of $h$ in the above expansion is identified with the $\mathbb Z_2$ summation variable $g=0,1$ in $\frac{1}{2}\sum_g (-1)^{gQ}$ implementing the corresponding  orbifold projection to invariant states (\emph{i.e.} to states with even eigenvalue $Q=0\,(\text{mod}\,2)$ under the $\mathbb Z_2$ action). For every twisted $\chi^I$ direction, the invariance of the worldsheet supercurrent requires a similar twist in the normal ordered product $y^I \omega^I \sim \partial X^I$ (in other words, $y^I$ has to appear in $\beta'$ if $\chi^I$ does), corresponding to the $\mathbb Z_2$ rotation $X^I\to-X^I$ in the orbifold language. It is easy to check using \eqref{FFprojector} that this breaks the same  supersymmetries as the associated $\mathbb Z_2$ orbifold rotation. Indeed, $e^{i\pi \beta' \cdot F_S}$ is the phase picked up by the transformation of each state in the Ramond vacuum $|S\rangle$, which is either projected away or retained, depending on the choice of $C\left[S\atop \beta'\right]$. As before, also in this case the nature of the breaking is not apparent at this stage -- identifying the corresponding orbifold as freely acting or not depends on whether the twist $\beta'$ involves a translation of some internal coordinates, but a detailed discussion of this will be given in later sections.

Returning to the minimal example with two basis elements $\{\mathds{1},S\}$ discussed above, the expansion \eqref{Expansion1} splits the worldsheet fermions into the RNS fermions in $S$ and those in its complement. The latter fermions carry the boundary conditions dictated by $\mathds{1}+S$, and their contribution to the one-loop partition function is recognised as a Narain lattice of signature $(6,22)$ at the fermionic point achieved at special values for the internal metric, $B$-field and Wilson lines parametrising the $\text{SO}(6,22)/\text{SO}(6)\times\text{SO}(22)$ moduli space. To see this, use the expansion \eqref{Expansion1} to write the partition function as
\begin{equation}
	Z=\frac{1}{\eta^{12}\,\bar\eta^{24}}\,\frac{1}{2}\sum_{a,b=0,1} \frac{1}{2}\sum_{m,n=0,1} C\left[aS+m(\mathds{1}+S)\atop bS+n(\mathds{1}+S)\right]\,\Theta\left[a\atop b\right]^4\, \Theta\left[m\atop n\right]^6 \, \bar\Theta\left[m\atop n\right]^{22} \,.
	\label{PartitionExample}
\end{equation}
By repeated application of the relations \eqref{modularity}, one finds that the $(a,b)$ and $(m,n)$ spin structures decouple in the phase
\begin{equation}
	C\left[aS+m(\mathds{1}+S)\atop bS+n(\mathds{1}+S)\right] = (-1)^{a+b+ab\,\tilde{C}\left[S\atop S\right]} \, (-1)^{mn\,\tilde{C}\left[\mathds{1}+S\atop \mathds{1}+S\right]} \,,
	\label{PhaseExample}
\end{equation}
where the new coefficients $\tilde C\in\{0,1\}$, defined by $C\left[\alpha\atop\beta\right] = (-1)^{\tilde{C}\left[\alpha\atop\beta\right]}$, shall be henceforth referred to as the \emph{associated GGSO coefficients}. Plugging \eqref{PhaseExample} into \eqref{PartitionExample} yields
\begin{equation}
	Z=\frac{1}{\eta^{12}\,\bar\eta^{24}} \left\{ \frac{1}{2}\sum_{a,b=0,1} (-1)^{a+b+ab\,\tilde{C}\left[S\atop S\right]} \,\Theta\left[a\atop b\right]^4 \right\} \left\{ \frac{1}{2}\sum_{m,n=0,1} (-1)^{mn\,\tilde{C}\left[\mathds{1}+S\atop \mathds{1}+S\right]}\, \Theta\left[m\atop n\right]^6 \, \bar\Theta\left[m\atop n\right]^{22} \right\} \,,
	\label{PartitionExample2}
\end{equation}
which shows that the partition function manifestly factorises into the RNS part (first curly brackets) and a (6,22) lattice sum realised by the remaining free fermions. There are only two independent phases left arbitrary by the modularity conditions, $C\left[S\atop S\right]$ and $C\left[\mathds{1}+S\atop \mathds{1}+S\right]$, both of which can be seen to correspond to choices of chirality conventions for the spinorial representations entering the characters of $\text{SO}(8)$, $\text{SO}(6)$ and $\text{SO}(22)$.
It is straightforward to cast the lattice sum into its Narain form by substituting the sum representation \eqref{MumfordTheta} of the theta functions into the second curly brackets in \eqref{PartitionExample2}, constructing the lattice vectors, recognising that $m=0,1$ produce representations with integer and half-integer charges, respectively, and finally performing the sum over $n=0,1$ which imposes self-duality. By comparison with the general form of the Narain lattice, one may then read off the fermionic point values for the internal metric $G_{IJ}$, Kalb-Ramond field $B_{IJ}$, and Wilson lines $Y_I^a$, but such an analysis is not particularly illuminating for our purposes and will not be performed here.

Let us briefly pause here to reflect on our steps and what they achieved. We started with the simplest free fermionic model admitting only two basis vectors $\mathscr{B}=\{\mathds{1},S\}$, and a GGSO matrix with 2 independent entries $C\left[\mathds{1}\atop\mathds{1}\right]$ and $C\left[S\atop\mathds{1}\right]$, with all others being determined by the modularity conditions \eqref{modularity}. We then picked an equivalent basis $\mathscr{B}'=\{S,\mathds{1}+S\}$, such that the new basis vectors had trivial overlap, \emph{i.e.} $S\cap (\mathds{1}+S)=\varnothing$. The general boundary condition vectors $\alpha,\beta\in \Xi$ associated to the spin structure $\left[\alpha\atop\beta\right]$ were then expanded in the new basis
\begin{equation}
	\alpha = aS + m(\mathds{1}+S)  \ ,\quad \beta = bS+n(\mathds{1}+S) ~~(\text{mod}\,2)\,.
	\label{Expansion3}
\end{equation}
The pairs of $\mathbb Z_2$ parameters $(a,b)$ and $(m,n)$ are conjugate to the two basis vectors of $\mathscr{B}'$ and encode the boundary condition assignments of the fermions in each basis element (viewed as a set).
Importantly, the expansion \eqref{Expansion3}  partitions the set of all worldsheet fermions into two subsets, the elements of which share the same boundary conditions. Specifically, the first subset comprises the left-moving RNS fermions in $S$, which carry the common boundary conditions $\left[ a\atop b\right]$, while the second set contains all other fermions -- the latter carry the common boundary conditions $\left[ m\atop n\right]$. The one-loop partition function \eqref{PartitionExample} then took the form of products of theta functions with characteristics $\left[ a\atop b\right]$ and $\left[ m\atop n\right]$, reflecting the aforementioned partition of the fermions. The product of theta functions is then summed over the independent parameters $a,b,m,n=0,1$, after being multiplied by  the phase \eqref{PhaseExample} involving the summation variables $a,b,m,n$, and two independent GGSO coefficients which were chosen to be $C\left[S\atop S\right]$ and $C\left[\mathds{1}+S\atop \mathds{1}+S\right]$. Note that the knowledge of $\Xi$ alone (or, equivalently, the basis) fixes the form of the partition function 
\begin{equation}
	Z = \frac{1}{\eta^{12}\,\bar\eta^{24}}\,\frac{1}{2^2}\sum_{a,m =0,1\atop b,n=0,1} (-1)^{I\left[a\,,\,m\atop b\,,\,n\right]} \, \,\Theta\left[a\atop b\right]^4\, \Theta\left[m\atop n\right]^6 \, \bar\Theta\left[m\atop n\right]^{22} \,,
	\label{PartitionExample3}
\end{equation}
up to a phase $(-1)^I$ that depends on the choice of the GGSO matrix. In the previous discussion, $I\left[a\,,\,m\atop b\,,\,n\right]$ was fixed in \eqref{PhaseExample} by repeatedly applying the modularity constraints \eqref{modularity}. Such a direct approach, however, becomes extremely tedious as soon as more basis vectors are introduced. For semi-realistic models this would be practically hopeless. 

A much more efficient approach would be to impose the modular invariance and factorisation conditions directly in the partition function of the form \eqref{PartitionExample3}, and derive the analogue of \eqref{modularity} directly in terms of the $(a,b)$ and $(m,n)$ parameters, rather than in terms of the vectors $\alpha,\beta\in \Xi$. Defining the (column) vectors $\textbf{a}=(a,m)^T$ and $\textbf{b}=(b,n)^T$ allows us to collectively denote the boundary condition assignment as $\left[a\,,\,m\atop b\,,\,n\right]\equiv \left[\textbf{a}\atop\textbf{b}\right]$. For the example at hand, the modular invariance conditions for the $I$'s can be directly extracted\footnote{In general, $C\left[\beta\atop -\alpha\right]=C\left[\beta\atop\alpha\right]^*$, so that in the case of purely periodic or anti-periodic boundary conditions we have simply $I\left[\textbf{b}\atop -\textbf{a}\right] = -I\left[\textbf{b}\atop\textbf{a}\right]= +I\left[\textbf{b}\atop\textbf{a}\right]$ (mod 2).} from \eqref{modularity} and take the form
\begin{equation}
	\begin{split}
		I\left[\textbf{a} \atop \textbf{b}\right] &= I\left[\textbf{a} \atop \textbf{b}-\textbf{a}+\mathds{1}\right] + a^2+1  ~~(\text{mod}\,2)\,,\\
		I\left[\textbf{a}  \atop \textbf{b} \right] &= I\left[\textbf{b} \atop \textbf{a}\right]   ~~(\text{mod}\,2)\,,\\
		I\left[\textbf{a} \atop \textbf{b}+\textbf{b}'\right] &= I\left[\textbf{a} \atop \textbf{b}\right] + I\left[\textbf{a} \atop \textbf{b}'\right] + a ~~(\text{mod}\,2)\,,\\
	\end{split}
	\label{ExampleModularity}
\end{equation}
The first two are simply the conditions of one-loop modular invariance, namely the invariance under $\tau\to\tau+1$ and $\tau\to-1/\tau$, respectively. The third line is the two-loop condition arising from factorisation and implies that $I$ can be at most linear in each lower characteristic. The second condition, imposing symmetry under the exchange $\textbf{a}\leftrightarrow\textbf{b}$, together with the third, then imply that the solution can be cast in the form
\begin{equation}
	I\left[\textbf{a} \atop \textbf{b}\right]  = a+b+ \textbf{a}^T \textbf{M} \textbf{b} ~~(\text{mod}\,2)\,,
	\label{ExampleAnsatz}
\end{equation}
for some $2\times 2$ symmetric matrix $\textbf{M}$ with integer entries (defined modulo 2). We have yet to impose the first condition in \eqref{ExampleModularity}. Doing so further constrains the matrix elements $M_{ij} \equiv(\textbf{M})_{ij}$ to satisfy the relations
\begin{equation}
	M_{ii} = M_{i1}+M_{i2} ~~(\text{mod}\,2)\,,
\end{equation}
which make the matrix diagonal, $M_{12}=M_{21}=0$, and result in the (global) factorisation of the RNS fermion contributions to the one-loop partition function from the Narain lattice. We are, hence, left with two independent entries $M_{11}$ and $M_{22}$, either of which may take the values 0 or 1, and the phase in the partition function \eqref{PartitionExample3} takes the form
\begin{equation}
	(-1)^{I\left[a\,,\,m\atop b\,,\,n\right]} = (-1)^{a+b+ab M_{11}}\,(-1)^{mn M_{22}}\,.
	\label{ExampleIntermediate}
\end{equation}
Obviously, a direct comparison with \eqref{PhaseExample} would fix these matrix elements in terms of the associated GGSO coefficients.
However, it is important to see that we can independently obtain this identification without having to resort to \eqref{PhaseExample}. First, note that the two basis elements $S$ and $\mathds{1}+S$ correspond to the unit vectors $\textbf{a}=(1,0)^T\equiv\textbf{e}_1$ and $\textbf{a}=(0,1)^T\equiv \textbf{e}_2$, respectively (and, similarly, for $\textbf{b}$). This bijection allows one to solve \eqref{ExampleAnsatz} for the matrix elements $M_{ij}$, simply by evaluating \eqref{ExampleAnsatz} at $\textbf{a}=\textbf{e}_i$ and $\textbf{b}=\textbf{e}_j$,
\begin{equation}
	M_{ij} = \left.\left\{a+b+\tilde C\left[aS +m(\mathds{1}+S)\atop bS+n(\mathds{1}+S)\right]\right\}\right|_{\textbf{a}=\textbf{e}_i\atop \textbf{b}=\textbf{e}_j}\,.
\end{equation}
From this, we immediately recover $M_{11} = \tilde C\left[S\atop S\right]$ and $M_{22} = \tilde C\left[\mathds{1}+S\atop \mathds{1}+S\right]$, in exact agreement with \eqref{PhaseExample}.
At this stage, all that is left is to recognise the Narain lattice in \eqref{PartitionExample2} and, if needed, deform the theory away from the fermionic point. In this case, the resulting partition function would still be given by \eqref{PartitionExample2}, except for the fact that the contributions inside the second curly bracket would be replaced by the general Narain lattice sum $\Gamma_{6,22}(\tau,\bar\tau;G,B,Y)$, as a function of the moduli.

Before ending this section, let us briefly comment on the modular properties of the solution \eqref{ExampleAnsatz}. The one-loop modular transformations $\tau\to\tau+1$ and $\tau\to-1/\tau$, induce an action on the space of characteristics of the form $\left[\textbf{a}\atop\textbf{b}\right]\to \left[\textbf{a}\atop \textbf{b}-\textbf{a}+\mathds{1}\right]$ and $\left[\textbf{a}\atop\textbf{b}\right]\to \left[\textbf{b}\atop\textbf{a}\right]$, respectively. Note that the spin-statistics part, $a+b$, in the phase exponent \eqref{ExampleAnsatz} is clearly symmetric under $\textbf{a}\leftrightarrow\textbf{b}$, and therefore, it is invariant under inversions, $\tau\to-1/\tau$. However, it is not invariant under translations $\tau\to\tau+1$, but this is needed to precisely cancel the additional phases arising from the transformation of the Dedekind functions and the $\Theta\left[a\atop b\right]^4$ factor in \eqref{PartitionExample3}. The second term $\textbf{a}^T\textbf{M}\textbf{b}$ in \eqref{ExampleAnsatz} is, instead, modular invariant and was seen to reflect the freedom to pick $N(N-1)/2+1=2$ independent GGSO phases in the free fermionic framework (applicable to $N=2$ basis elements). This splitting of the phase into a ``modular balancing" part and a modular invariant part is actually general. The non-trivially transforming phase is dictated by the choice of basis and so, in practice, it will be chosen from the beginning to cancel phases arising in the transformation of Theta and Dedekind functions, while absorbing any remaining freedom (corresponding to GGSO choices) into a modular invariant phase. It is not difficult to see that, in the specific example considered here, $(-1)^{ab}$ and $(-1)^{mn}$ are the only modular invariant phases of the form $(-1)^{\textbf{a}^T\textbf{M}\textbf{b}}$ that can be constructed out of $(a,b)$ and $(m,n)$. Whether or not to include them into the partition function is a choice parametrised by $M_{11}$ and $M_{22}$, and we again recover \eqref{ExampleIntermediate}.

This completes the map of the minimal free fermionic theory generated by $\mathscr{B}=\{\mathds{1},S\}$ to its equivalent toroidal compactification. Of course, in this particular example the map itself was rather trivial. Nevertheless, it served as an illustration of the salient features entering the general map that will be developed in the following sections. Indeed, the logic of organising the basis vectors such that they partition the fermions into non-intersecting sets (with fermions in each set sharing common boundary conditions), the subsequent interpretation of additional basis vectors as twists of their boundary conditions, the idea of packaging the latter in the one-loop partition function in terms of the components $(a,m)$, $(b,n)$ of vectors $\alpha,\beta\in \Xi$ on which to impose modularity constraints, and solving them in terms of a phase similar to \eqref{ExampleAnsatz} that is itself determined by the independent GGSO coefficients, is not particular to the simple example considered above. Rather, it is the basis for making the correspondence between semi-realistic free fermionic and orbifold theories manifest. 


\section{Generalising the map}
\label{GenericMap}

In the previous section we highlighted the main ingredients of the map in the simple setup of the minimal (toroidal) basis $\{\mathds{1},S\}$. We are now ready to move on to the general case.
Our starting point is a consistent free-fermionic theory defined in terms of a basis $\mathscr{B}=\{v_i\}$ with $i=1,\ldots,N$, and an $N\times N$, phase-valued GGSO matrix $C_{ij}=C\left[v_i \atop v_j\right]$.
We are interested in the case where $S\in\Xi$, so without loss of generality, we take $S\in\mathscr{B}$.

As before, our first task is to define an equivalent basis $\mathscr{B}' = \mathscr{P}\cup\mathscr{T}$, such that vectors in $\mathscr{P}$ define a partition of the set of all fermions into equivalence classes, with fermions in each class sharing the same boundary conditions, while vectors in $\mathscr{T}$ generate twists of those boundary conditions. Importantly, the reduced additive group $\Xi' = \text{span}(\mathscr{P})$ must itself generate a consistent free fermionic theory, corresponding to the original theory before the orbifold twists are introduced. In particular, this implies $\mathds{1}\in\text{span}(\mathscr{P})$. In the toroidal orbifold formulation, the RNS fermions initially share common boundary conditions $\left[a\atop b\right]$, which are then twisted by the orbifold action. To match with the orbifold representation, we therefore pick $S\in\mathscr{P}$. For the same reason, basis vectors $t\in\mathscr{B}'$ acting non-trivially on the RNS fermions (\emph{i.e.} on the elements of $S$) will necessarily be elements of $\mathscr{T}$. 

Despite these requirements, there may still be some residual arbitrariness in how $\mathscr{B}'$ splits into `pure boundary condition' vectors in $\mathscr{P}$, and twists in $\mathscr{T}$. For example, consider the 16 complex right-moving fermions $\bar\psi^{1},\ldots,\bar\psi^5,\bar\eta^1,\bar\eta^2,\bar\eta^3,\bar\phi^1,\ldots,\bar\phi^8$, that we will often refer to as the `Kac-Moody fermions'. In ten dimensions, assigning common boundary conditions to all 16 of them leads to the $\text{SO}(32)$ theory, while partitioning them into two sets of 8 and assigning separate boundary conditions to each set yields the $\text{E}_8\times\text{E}_8$ theory. However, we could alternatively obtain it starting from $\text{SO}(32)$, by introducing a $\mathbb{Z}_2$ orbifold that twists the boundary conditions of 8 fermions
\begin{equation}
	\frac{1}{2}\sum_{h,g=0,1}\frac{1}{2}\sum_{k,\ell=0,1} \bar\Theta\left[k \atop \ell\right]^8 \, \bar\Theta\left[k+h \atop \ell+g\right]^8 = \frac{1}{2}\sum_{k,\ell=0,1}\bar\Theta\left[k \atop \ell\right]^8 \,\frac{1}{2}\sum_{\rho,\sigma=0,1}\bar\Theta\left[\rho \atop \sigma\right]^8 \,.
	\label{SO32E8}
\end{equation}
Here, the l.h.s. corresponds to the right-moving contribution of the $\text{SO}(32)$ theory with all 16 fermions sharing the boundary conditions $\left[k\atop\ell\right]$ before the twist, while $\left[h\atop g\right]$ implements the $\mathbb Z_2$ twist of the last 8 fermions. In ten dimensions the auxiliary fermions $y,\omega$ are not present and, hence, the l.h.s. corresponds to the choice $\mathscr{P}=\{S,\mathds{1}+S\}$ and $\mathscr{T}=\{z\}$, with $\mathds{1}+S=\{\bar\psi^{1},\ldots,\bar\psi^5,\bar\eta^1,\bar\eta^2,\bar\eta^3,\bar\phi^1,\ldots,\bar\phi^8\}$ and $z=\{\bar\phi^1,\ldots,\bar\phi^8\}$.
The r.h.s. of \eqref{SO32E8} corresponds instead to the case where the Kac-Moody fermions are split into two sets, and are assigned the independent boundary conditions $\left[k\atop\ell\right]$, $\left[\rho\atop\sigma\right]$, respectively. Here, there is no twist, $\mathscr{T}=\varnothing$, and the $\text{E}_8\times\text{E}_8$ theory is constructed directly by the `pure' boundary conditions basis, $\mathscr{P}=\{S,\mathds{1}+S+z,z\}$. 

In constructing the map to orbifolds, we will often find it convenient to pick $\mathscr{P}$ such that $\left[a\atop b\right]$ describes the boundary conditions of the RNS fermions, $\left[k\atop\ell\right]$ describes the Kac-Moody fermions associated to the first $\text{E}_8$ factor, $\left[\rho\atop\sigma\right]$ are the boundary conditions of fermions making up the second $\text{E}_8$ factor, while using $\left[\gamma\atop\delta\right]$ (possibly with indices) to denote those of auxiliary fermions arising from compactification on $T^6$. Similarly, elements of $\mathscr{T}$ will generate $\mathbb{Z}_2$ twists of these boundary conditions and we adopt a notation where $\left[h\atop g\right]$ refers to rotations, $\left[H_I\atop G_I\right]$ refers to translations, and we reserve the symbols $\left[H\atop G\right]$, $\left[H'\atop G'\right], \ldots$, for twists of the Kac-Moody fermions which do not involve $T^6$ rotations. This notation will come handy in explicitly constructing the map for particular models. 

The expansion of the generic elements $\alpha,\beta\in\Xi$ in terms of the basis $\mathscr{B}' =\{v_i'\}$ takes the form
\begin{equation}
	\alpha = A_i v_i' \ , \quad \beta = B_i v_i' \,,
	\label{BetaPrimeExpansion}
\end{equation}
where the components $A_i$, $B_i$ of $\alpha,\beta$, respectively, are defined modulo 2. We will see below that they (partly) correspond to the independent summation parameters in the orbifold representation. We can think of $A_i, B_i$ as $N$-dimensional column vectors of the form
\begin{equation}
	\textbf{A}=(a,k,\rho,\gamma_1,\ldots;h_1,h_2,\ldots)^T \ ,\quad \textbf{B} = (b,\ell,\sigma,\delta_1, \ldots; g_1,g_2,\ldots)^T \,,
	\label{ComponBprime}
\end{equation}
where $a,b$ are the components associated to $v_1'=S$, and the semicolons separate the pure boundary conditions of $\mathscr{P}$ from the twists in $\mathscr{T}$. Clearly, the expansion will need to be suitably adapted to the problem (and basis) at hand, but the form displayed above will be sufficiently illustrative for the purpose of this discussion. 

Plugging this expansion into \eqref{FFpartition}, the GGSO coefficients become $C\left[A_i v_i' \atop B_j v_j'\right]$, and the summation is over all values of the components $A_i,B_i=0,1$. In this form, the characteristics of theta functions become linear combinations of the components $A_i$, $B_i$. More specifically, the characteristic of each theta function is either of the form $\left[ p \atop q\right]$ with $(p,q)$ being a component pair corresponding to pure boundary conditions $\mathscr{P}$, or of the twisted form $\left[ p + t \atop q + s\right]$ where $(t,s)$ is a linear combination of twist components in $\mathscr{T}$. Clearly, twist components never appear in theta characteristics on their own, but may only appear additively as offsets of the pure boundary conditions. This implies that they transform differently under genus-1 modular transformations. For example, under $\tau\to\tau+1$,
\begin{equation}
	\begin{pmatrix}
		p \\
		 q
	\end{pmatrix} \to \begin{pmatrix}
					p \\
					 q+p-1
				\end{pmatrix} \ ,\quad \begin{pmatrix}
									t \\
									s
								\end{pmatrix} \to \begin{pmatrix}
									t \\
									s+t
									\end{pmatrix} \,,
\end{equation}
while under $\tau\to-1/\tau$,
\begin{equation}
	\begin{pmatrix}
		p \\
		q 
	\end{pmatrix} \to \begin{pmatrix}
					q \\
					p
				\end{pmatrix} \ ,\quad \begin{pmatrix}
									t\\
									s
								\end{pmatrix} \to \begin{pmatrix}
											s \\
											t
											\end{pmatrix} \,.
\end{equation}

With theta function factors arranged in the same order that the corresponding fermions enter the element $\mathds{1}\in\Xi$ (using complex fermions whenever possible), we may now assemble together their upper and lower characteristics corresponding to a spin structure $\alpha,\beta\in\Xi$ into column vectors $\hat{\textbf{a}}$ and $\hat{\textbf{b}}$, respectively. Their $f$-th components in free fermion-space (\emph{i.e.} ascribed to the free fermion $f$) read
\begin{equation}
	(\hat{\textbf{a}})_f \equiv \alpha(f) = A_i v'_i(f) \ ,\quad (\hat{\textbf{b}})_f \equiv \beta(f) = B_i v'_i(f) \,.
\end{equation}
For example, in a typical $T^6/\mathbb Z_2\times \mathbb Z_2$ orbifold rotating the first two 2-torii by $(h_1,g_1)$ and the last two 2-torii by $(h_2,g_2)$, the first few components of these vectors (corresponding to the complexified fermions in $S$) would take the standard form
\begin{equation}
	\hat{\textbf{a}} = (a,a+h_1,a+h_2,a-h_1-h_2,\ldots)^T \ ,\quad \hat{\textbf{b}} = (b,b+g_1,b+g_2,b-g_1-g_2,\ldots)^T \,.
	\label{thetaArguments}
\end{equation}
The minus signs in the above expression are conventionally chosen to keep the orbifold representation and the Riemann identities simple.
Viewed as column vectors in free-fermion space, the only difference between the vectors $\hat{\textbf{a}},\hat{\textbf{b}}$ and the corresponding boundary condition vectors $\alpha,\beta$ defined in \eqref{DefAlpha} is that the latter have entries restricted to lie in the interval $(-1,1]$, whereas the former do not necessarily have this property. Nevertheless, the components of $\hat{\textbf{a}},\hat{\textbf{b}}$ may always be brought into this interval by adding suitable even integers. We shall denote by $[\hat{\textbf{a}}]$ the reduced representative of $\hat{\textbf{a}}$, such that all its entries lie in $(-1,1]$. It will be useful to define also the `reduction vector' $r(\hat{\textbf{a}})= (\hat{\textbf{a}}-[\hat{\textbf{a}}])/2$ as half the difference between the vector and its reduced representative. Clearly, $r(\hat{\textbf{a}})$ has integer components, and similarly for $r(\hat{\textbf{b}})$. Furthermore, it was proven in \cite{Antoniadis:1987wp} that the free fermionic construction rules necessarily imply the integrality of dot products of reduction vectors, namely, $r(\hat{\textbf{a}})\cdot r(\hat{\textbf{b}})\in\mathbb{Z}$.

In terms of $\hat{\textbf{a}}, \hat{\textbf{b}}$, we can rewrite the partition function corresponding to \eqref{FFpartition} in the form
\begin{equation}
	Z = \frac{1}{\eta^{12}\,\bar\eta^{24}}\,\frac{1}{2^N}\sum_{\{A\},\{B\}} e^{i\pi\, \tilde C\left[ [\hat{\textbf{a}}] \atop [\hat{\textbf{b}}] \right] } \, \Theta\left[ [\hat{\textbf{a}}] \atop [\hat{\textbf{b}}] \right] \,,
	\label{FFIntermediate}
\end{equation}
where the products of holomorphic and anti-holomorphic theta functions are now collectively denoted by $\Theta\left[ [\hat{\textbf{a}}] \atop [\hat{\textbf{b}}] \right]$. Notice that the characteristics are written in terms of the reduced representatives, since, viewed as column vectors in free-fermion space, $\alpha=[\hat{\textbf{a}}]$ and $\beta=[\hat{\textbf{b}}]$. The symbol $\{A\},\{B\}$ refers to the fact that we are summing all components $A_i, B_i = 0,1$ .  

Our strategy is to massage this expression into a conventional ``orbifold'' form in which the (twisted/shifted) Narain lattices can be most easily recognised. To do this, we must first take into account the fact that the theta functions in a conventional orbifold representation contain unreduced characteristics of the form \eqref{thetaArguments}, whereas those in \eqref{FFIntermediate} start in reduced form by construction. Indeed, theta functions are periodic in their characteristics (with period 2), modulo multiplicative phase factors that we need to keep track of. This is straightforward to do, if we notice that \eqref{MumfordTheta} implies
\begin{equation}
	\Theta\left[ \phi\atop\chi \right] (z;\tau) = e^{-i\pi \,r(\phi) \chi} \, \Theta\left[ [\phi] \atop [\chi]\right](z;\tau) \,.
\end{equation}
Furthermore, we switch to the standard convention $\Theta\to\vartheta$ of theta functions appearing in much of the orbifold litterature, related to \eqref{MumfordTheta} by 
\begin{equation}
	\vartheta \left[ \phi\atop\chi\right](z;\tau) = e^{i\pi\phi\chi/2} \,\Theta\left[\phi\atop\chi\right] (z;\tau) \,.
	\label{thetaConvention}
\end{equation}
With these changes, the partition function takes the form
\begin{equation}
	Z = \frac{1}{\eta^{12}\,\bar\eta^{24}}\,\frac{1}{2^N}\sum_{\{A\},\{B\}} e^{i\pi\, \Phi\left[ \hat{\textbf{a}} \atop \hat{\textbf{b}} \right] } \, \vartheta\left[ \hat{\textbf{a}} \atop \hat{\textbf{b}} \right] \,,
	\label{FFIntermediate2}
\end{equation}
where $\vartheta\left[ \hat{\textbf{a}} \atop \hat{\textbf{b}} \right]$ collectively denotes the products of holomorphic and anti-holomorphic theta functions in the convention \eqref{thetaConvention}, and
the phase $\Phi$ is determined by the associated GGSO coefficients,
\begin{equation}
	\Phi\left[ \hat{\textbf{a}} \atop \hat{\textbf{b}} \right] = \tilde C\left[ [\hat{\textbf{a}}] \atop [\hat{\textbf{b}}] \right] + r(\hat{\textbf{a}})\cdot \hat{\textbf{b}} -\frac{1}{2}\hat{\textbf{a}}\cdot\hat{\textbf{b}} ~~ (\text{mod}\ 2)\,.
	\label{PhaseRelation}
\end{equation}
Using the fermionic construction rules \eqref{FFrules}, it is not difficult to see that the last term, $-\frac{1}{2}\hat{\textbf{a}}\cdot\hat{\textbf{b}}$, is always an integer. In fact, it vanishes (modulo 2) for symmetric orbifolds with standard embedding whose rotational action preserves spacetime supersymmetry, provided the signs of the twists are chosen to sum up to zero, \emph{e.g.} as in \eqref{thetaArguments}. 

Plugging \eqref{PhaseRelation} for $\Phi$ into the modularity relations \eqref{modularity} gives rise to equivalent conditions in terms of the $\Phi$'s. These will now solved as follows. We first decompose
\begin{equation}
	\Phi\left[ \hat{\textbf{a}} \atop \hat{\textbf{b}} \right] =\Lambda\left[ \hat{\textbf{a}} \atop \hat{\textbf{b}} \right] +\Omega\left[ \hat{\textbf{a}} \atop \hat{\textbf{b}} \right] \,,
\end{equation}
into a modular invariant part $\Omega$, satisfying
\begin{equation}
	\begin{split}
		\Omega\left[ \hat{\textbf{a}} \atop \hat{\textbf{b}} \right] &= \Omega\left[ \hat{\textbf{a}} \atop \hat{\textbf{b}} - \hat{\textbf{a}} + \mathds{1} \right]  ~~ (\text{mod}\ 2) \,,\\
		\Omega\left[ \hat{\textbf{a}} \atop \hat{\textbf{b}} \right] &= -\Omega\left[ \hat{\textbf{b}} \atop \hat{\textbf{a}} \right]  ~~ (\text{mod}\ 2) \,,\\
		\Omega\left[ \hat{\textbf{a}} \atop \hat{\textbf{b}}+\hat{\textbf{b}}' \right] &= \Omega\left[ \hat{\textbf{a}} \atop \hat{\textbf{b}} \right]+\Omega\left[ \hat{\textbf{a}} \atop \hat{\textbf{b}}' \right]  ~~ (\text{mod}\ 2) \,,
	\end{split}
	\label{ModularityOmega}
\end{equation}
and a `modular balancing' part $\Lambda$, transforming non-trivially as
\begin{equation}
	\begin{split}
		\Lambda\left[ \hat{\textbf{a}} \atop \hat{\textbf{b}} \right] &= \Lambda\left[ \hat{\textbf{a}} \atop \hat{\textbf{b}} - \hat{\textbf{a}} + \mathds{1} \right] +1+\frac{1}{4}\,\hat{\textbf{a}}\cdot \hat{\textbf{a}} ~~ (\text{mod}\ 2) \,,\\
		\Lambda\left[ \hat{\textbf{a}} \atop \hat{\textbf{b}} \right] &= -\Lambda\left[ \hat{\textbf{b}} \atop \hat{\textbf{a}} \right] -\frac{1}{2}\hat{\textbf{a}}\cdot\hat{\textbf{b}} ~~ (\text{mod}\ 2) \,,\\
		\Lambda\left[ \hat{\textbf{a}} \atop \hat{\textbf{b}}+\hat{\textbf{b}}' \right] &= \Lambda\left[ \hat{\textbf{a}} \atop \hat{\textbf{b}} \right]+\Lambda\left[ \hat{\textbf{a}} \atop \hat{\textbf{b}}' \right] + a ~~ (\text{mod}\ 2) \,.
	\end{split}
	\label{ModularityLambda}
\end{equation}
As with \eqref{ExampleModularity}, also here the transposition and factorisation conditions in the second and third line of \eqref{ModularityOmega}, require that $\Omega$ be (anti)symmetric\footnote{For $\mathbb Z_2$ boundary conditions, also the diagonal elements are allowed so that one should take $\Omega$ to be symmetric.} (modulo 2) under the exchange $\hat{\textbf{a}} \leftrightarrow \hat{\textbf{b}}$ and that it be bi-linear in the upper and lower parameters. We can therefore write 
\begin{equation}
	\Omega\left[ \hat{\textbf{a}} \atop \hat{\textbf{b}} \right] = \textbf{A}^T \,\textbf{M} \,\textbf{B} ~~(\text{mod}\ 2)\,,
	\label{OmegaSolution}
\end{equation}
in terms of an $N\times N$ (anti)symmetric matrix  $\textbf{M}$, defined modulo 2. The first condition in \eqref{ModularityOmega} imposes the additional constraint
\begin{equation}
	M_{ii} = \sum_{j=1}^{N} M_{ij}\nu_j ~~(\text{mod}\ 2)\,,
	\label{Mcond}
\end{equation}
for $i=1,\ldots,N$, where $\nu_j\equiv \nu(v_j')$ is defined to be zero if $v_j'\in\mathscr{T}$, and one if $v_j' \in\mathscr{P}$. The matrix $\textbf{M}$ represents the freedom in including modular invariant phases and represents the corresponding freedom in choosing the GGSO coefficients. As a check, in the case of purely real boundary conditions, there are $N(N-1)/2$ conditions from the (anti)symmetry of $\textbf{M}$ but only $N-1$ conditions from the modular $\tau\to\tau+1$ transformation, since summing \eqref{Mcond} over all $i$ in the non-twist directions trivially vanishes (modulo 2). As a result, there are $N(N-1)/2+1$ independent elements in $\textbf{M}$, which precisely match the number of independent GGSO phases.

Let us now discuss the modular transforming contribution $\Lambda$. The presence of the additive term $a$ in the factorisation condition \eqref{ModularityLambda}, indicates that $\Lambda$ must contain the term $a+b$, giving the correct spin-statistics connection, as expected. Furthermore, the last two terms in the first condition of \eqref{ModularityLambda} are partly cancelled by the transformation of the same spin statistics term. Indeed, the term $\frac{1}{4}\,\hat{\textbf{a}}\cdot \hat{\textbf{a}} = a^2 + \ldots$ (mod 2) effectively picks the components of the expansion of $\alpha\in \Xi$ on $\mathscr{B}'$ corresponding to basis vectors defining the boundary conditions (or twists) of 4 complex (equiv. 8 real) fermions. Together with the factorisation condition, this implies that $\Lambda\left[ \hat{\textbf{a}} \atop \hat{\textbf{b}} \right]=a+b+\ldots$ contains exactly one strictly linear term: the spin-statistics combination $a+b$;  the ellipses denote other contributions required to cancel any remaining $a$-independent terms in $\frac{1}{4}\,\hat{\textbf{a}}\cdot \hat{\textbf{a}}$. If present, they must necessarily\footnote{Strictly linear combinations other than $a+b$ would violate the factorisation condition \eqref{ModularityLambda} (and spin-statistics). Bi-linears of pure boundary condition components in $\mathscr{P}$ (such as $ab$) are modular invariant, so they are contained in $\Omega$ instead.} be symmetric bi-linears in twist-like variables\footnote{This does not mean that they are necessarily made out of the twist variables of $\mathscr{T}$ but, rather, in terms of combinations that transform as such. For instance, if $\mathscr{P}$ includes the $\mathbb{Z}_2$-valued pure boundary condition components $\left[p_1\atop q_1\right]$ and $\left[p_2 \atop q_2\right]$, then a contribution $\frac{1}{4}\,\hat{\textbf{a}}\cdot \hat{\textbf{a}} = a^2-p_1^2-p_2^2$ can be cancelled by setting $\Lambda = a+b+(p_1+p_2)(q_1+q_2)$.} $\left(t\atop s\right)$, \emph{i.e.} they are of the form $ts$. Strictly speaking, the modular balancing part is not uniquely defined, since for any $\Lambda$ satisfying \eqref{ModularityLambda} we can trivially generate another, simply by adding modular invariant contributions of the type \eqref{OmegaSolution}. Of course, this essentially reflects the arbitrariness of the splitting $\Phi=\Lambda+\Omega$, but the point is that we can always pick one choice of $\Lambda$, since the remaining arbitrariness will be absorbed into $\Omega$.

Upon fixing $\Lambda$ and absorbing any remaining arbitrariness into the modular invariant phase $\Omega$, equation \eqref{PhaseRelation} may now be used to fix the matrix elements $M_{ij}$ in terms of the associated GGSO coefficients. This is precisely what we aim to do in the remainder of this section. Although most of the analysis so far relied on the basis $\mathscr{B}'$ which made the distinction between pure boundary conditions and twists manifest, in practical applications the free fermionic theory is usually defined in terms of some other initial basis $\mathscr{B}$. In particular, the GGSO matrix $C_{ij}=C\left[v_i\atop v_j\right]$ is also given in terms of $\mathscr{B}$. Take an arbitrary vector $\alpha \in\Xi$ and expand it in the original basis $\mathscr{B}=\{v_i\}$ as $\alpha = \lambda_i(\alpha)\,v_i$. Its components $\lambda_i(\alpha)$ are clearly linear combinations of the components $A_i$ of the expansion \eqref{BetaPrimeExpansion} in terms of $\mathscr{B}'$, 
\begin{equation}
	\lambda_i(\alpha) = \tilde{R}_{ij} A_j(\alpha) \,,
\end{equation}
for some $\alpha$-independent invertible matrix $\tilde{R}_{ij}$ that we may easily read off by comparing the expansions in the two bases. We may now interpret the $A_i$'s as the linear maps
\begin{equation}
	A_i(\alpha) = (\tilde{R}^{-1})_{ij}\,v^\star_j(\alpha)\,,
	\label{invMap}
\end{equation}
expanded in terms of the dual basis $\{v^\star_i\}$ and define also $R_{ij} \equiv (\tilde{R}^{-T})_{ij}$. In particular, \eqref{invMap} implies $A_j(v_i) = R_{ij}$. Having determined the matrix $R_{ij}$ encoding the change of basis allows the values of all components $a,b,k,\ell, \ldots,h,g\ldots$ to be uniquely fixed in terms of the original basis vectors $v_i,v_j$ entering the characteristics of the GGSO coefficients. For example, in the spin structure $\left[ v_i\atop v_j\right]$, we have
 \begin{equation}
	a\to A_1(v_i) = R_{i1} \,, \qquad b\to B_1(v_j) = R_{j1} \,, \qquad \ldots \,,
	\label{ComponentMap}
 \end{equation}
 while the other components are obtained in a similar way. Applying this identification to \eqref{OmegaSolution} for the basis elements, we obtain
\begin{equation}
	\Omega_{ij} \equiv \Omega\left[ v_i \atop v_j \right] = A_r(v_i) M_{rs} B_s(v_j) = (\textbf{R}\,\textbf{M}\,\textbf{R}^T)_{ij} \,.
\end{equation}
The choice of $\Lambda$ allows us to similarly work out the matrix elements 
\begin{equation}
	\Lambda_{ij}\equiv \Lambda\left[ v_i \atop v_j\right]=A_1(v_i)+B_1(v_j)+\ldots \,.
\end{equation}
We may now plug these results into \eqref{PhaseRelation} and solve for $\textbf{M}$. To this end, define
\begin{equation}
	Q_{ij} = \left[  r(\hat{\textbf{a}})\cdot \hat{\textbf{b}} -\frac{1}{2}\hat{\textbf{a}}\cdot\hat{\textbf{b}} \right]_{A_m(v_i)\to R_{im} \atop B_n(v_j)\to R_{jn} } - \Lambda_{ij} \,,
	\label{OffsetPhase}
\end{equation}
as an ``offset" matrix encoding the effect of the theta function periodicities. Special care should be given to the evaluation of the quantity in the square brackets of \eqref{OffsetPhase} because of the presence of the reduction vector $r(\hat{\textbf{a}})$. In particular, the Lorentzian product $r(\hat{\textbf{a}})\cdot \hat{\textbf{b}}$ should first be expanded in terms of the characteristics entering the theta functions \eqref{thetaArguments}. The components $A_m, B_n$ should then be replaced by the corresponding $R$-entries according to \eqref{ComponentMap}, while keeping track of periodicities that can contribute to the reduction vector. With these definitions in place, we may finally write the expression for
\begin{equation}
	\textbf{M} = \textbf{R}^{-1}\left(\tilde{\textbf{C}} +  \textbf{Q}\right)\textbf{R}^{-T} \,,
	\label{Msolution}
\end{equation}
in terms of the associated GGSO coefficients. This determines $\Omega$ and, therefore, also the phase $\Phi$ entering the partition function \eqref{FFIntermediate2}.

In a certain sense, we are halfway there. The partition function \eqref{FFIntermediate2}, organised in terms of the expansion \eqref{BetaPrimeExpansion}, contains theta functions with the characteristics \eqref{thetaArguments}, which are almost in standard orbifold form, at least as far as the rotational twists are concerned. Indeed, having separated the basis into the twists $\mathscr{T}$, and the pure boundary condition vectors $\mathscr{P}$ partitioning the space of free-fermions, the components $(\gamma,\delta)$ in $\mathscr{P}$ corresponding to directions orthogonal to the RNS sector $S$ and to the Kac-Moody fermions will necessarily provide the boundary condition variables associated to the auxiliary fermions $y^I,\omega^I$ and  their right-moving counterparts. In other words, $\left[\gamma\atop\delta\right]$ become the summation variables that, together with the corresponding theta functions, define the Narain lattice $\Gamma_{6,6}$ for the internal $T^6$, at the fermionic point. Of course, $\Gamma_{6,6}$ will be typically twisted by the rotational elements $(h,g)$ of $\mathscr{T}$ -- this will be recognised by the way the twists $(h,g)$ enter the characteristics of the theta functions associated to the auxiliary fermions. 

In general, the lattices may also be shifted, but this may not be directly obvious in the fermionic variables $\left[\gamma\atop\delta\right]$. On the one hand, in (rotationally) twisted sectors, momenta and windings are eliminated by the rotation and so, the triple product identity of theta functions $\vartheta_2\vartheta_3\vartheta_4=2\eta^3$ can be used to cast the corresponding lattice into its familiar orbifold form. On the other hand, in sectors where rotational twists are trivial, the momenta and windings are still present and one recovers a shifted lattice, which has to be brought into standard Narain form so that its moduli dependence may be recognised. This last step can be rather involved and amounts to replacing the theta functions corresponding to the auxiliary fermions by their sum representation, combining exponents, and performing the sums over $(\gamma,\delta)$. The shift orbifold labels $(H_a,G_a)$ will appear only after appropriately redefining the lattice summation variables. We defer the explicit details for the next section, where we will focus specifically on $T^6/\mathbb Z_2\times \mathbb Z_2$ orbifolds.


\section{Symmetric $T^6/\mathbb{Z}_2\times\mathbb{Z}_2$ with and without supersymmetry}
\label{SymExample}

Having outlined the general strategy of the map in Section \ref{GenericMap}, it will now be instructive to apply it to a specific class of semi-realistic models. We will work in the case where all fermions have boundary conditions $\alpha(f)=0,1$, defined by the symmetric basis $\mathscr{B}=\{v_i\}$ of \cite{Faraggi:2004rq}, containing the $12$ vectors 
\begin{equation}
	\begin{split}
		v_1 &= \mathds{1} \,,\\ 
		v_2 &= S = \{\psi^\mu, \chi^1,\ldots,\chi^6\} \,,\\
		v_3 &= \{y^1,\omega^1 ; \bar y^1, \bar\omega^1\} \,,\\
		v_4 &= \{y^2,\omega^2 ; \bar y^2, \bar\omega^2\} \,,\\
		      & \ldots		\\
		v_8 &= \{y^6,\omega^6 ; \bar y^6, \bar\omega^6\} \,,\\
		v_9 &= b_1 = \{ \chi^3, \ldots, \chi^6, y^3, \ldots, y^6 ; \bar y^3,\ldots,  \bar y^6, \bar\psi^1,\ldots, \bar\psi^5,\bar\eta^1\} \,,\\
  	    v_{10} &= b_2 = \{ \chi^1, \chi^2, \chi^5, \chi^6, y^1, y^2, y^5, y^6 ; \bar y^1, \bar y^2, \bar y^5, \bar y^6, \bar\psi^1,\ldots, \bar\psi^5,\bar\eta^2\} \,,\\	
	    v_{11} &= z_1 = \{\bar \phi^1,\ldots,\bar\phi^4\} \,,\\
	    v_{12} &= z_2 = \{\bar \phi^5,\ldots,\bar\phi^8\} \,.
	\end{split}
	\label{SymmetricBasis}
\end{equation}
As explained earlier, the above notation for basis vectors represents them as sets that include the fermions which are periodic in the given sector, though we shall alternatively view them as boundary condition vectors in the space of free fermions whenever required. There are 67 independent GGSO phases to be chosen, each choice giving rise to a consistent string theory (modulo stability issues in the non-supersymmetric cases, but this is beside the point here). 

It is now time to put the machinery discussed in Section \ref{GenericMap} into action. We first change the basis to $\mathscr{B}' =\mathscr{P}\cup\mathscr{T}$ by splitting up the original basis into pure boundary condition assignments $\mathscr{P}$ and twists $\mathscr{T}$. For $\mathscr{P}$, the natural choice partitioning the set $\mathds{1}$ into RNS fermions $S$, lattice (auxiliary) fermions, and the fermions associated to the two $\text{E}_8$'s is
\begin{equation}
	\mathscr{P} = \{ v_2, v_3, \ldots, v_8, v_1+\ldots+v_8+v_{11}+v_{12}, v_{11}+v_{12} \} \,.
\end{equation}
Note that we could have chosen to further partition the fermions of the second $\text{E}_8$ into two sets of four complex fermions, by including $v_{11}$ and $v_{12}$ as separate elements of $\mathscr{P}$. Here, we have opted to include instead only the combination $v_{11}+v_{12}$, which is more natural for making contact with orbifold compactifications of the $\text{E}_8\times\text{E}_8$ heterotic string. With this choice, the breaking of the second $\text{E}_8$ into two $\text{SO}(8)$ factors will be accomplished by suitably including $v_{12}$ in the twist basis. Indeed, $\mathscr{T}$ includes three twist elements
\begin{equation}
	\mathscr{T} = \{v_{10}, v_9, v_{12} \} \,.
\end{equation}
The first two elements, $v_{10}=b_2$ and $v_{9}=b_1$, twist the boundary conditions of $(\chi,y)$ pairs and, therefore, correspond to $\mathbb{Z}_2$ rotations in the corresponding directions. The class of theories under consideration will, hence, correspond to $T^6/\mathbb{Z}_2\times\mathbb{Z}_2$ orbifold compactifications. We next define the expansion components of generic vectors $\alpha,\beta\in\Xi$ in terms of the new basis $\mathscr{B}'$ as
\begin{equation}
	\begin{split}
		\alpha &= a v_2 + \gamma_1 v_3+\ldots+\gamma_6 v_8+ k(v_1+\ldots+v_8+v_{11}+v_{12}) + \rho (v_{11}+v_{12}) \\
				&+ h_1 (v_1+\ldots v_8+v_{11}+v_{12}+v_{10})+h_2 (v_1+\ldots+v_8+v_{11}+v_{12}+v_{9})+H v_{12}\,, \\
		\beta &= b v_2 + \delta_1 v_3+\ldots+\delta_6 v_8+ \ell(v_1+\ldots+v_8+v_{11}+v_{12}) + \sigma (v_{11}+v_{12}) \\
				&+ g_1 (v_1+\ldots v_8+v_{11}+v_{12}+v_{10})+g_2 (v_1+\ldots+v_8+v_{11}+v_{12}+v_{9})+G v_{12}\,.
	\end{split}
	\label{ModelExpandBprime}
\end{equation}
Assembling the components of $\alpha,\beta$ in terms of the basis $\mathscr{B}'$ into 12-dimensional column vectors, we define
\begin{equation}
		\textbf{A} = (a,\gamma_1,\ldots,\gamma_6, k,\rho ; h_1,h_2, H)^T\,,\quad \textbf{B} = (b,\delta_1,\ldots,\delta_6, \ell,\sigma ; g_1,g_2,G)^T \,,
		\label{BprimeBasisParams}
\end{equation}
where the semi-colon separates the pure boundary conditions from the twists, as in \eqref{ComponBprime}. We can now plug the expansion \eqref{ModelExpandBprime} into the partition function and organise it in the form \eqref{FFIntermediate2},
\begin{equation}
	\begin{split}
	Z&=\frac{1}{\eta^{12}\,\bar\eta^{24}}\frac{1}{2^{12}}\sum_{\{A\},\{B\}} (-1)^{\Phi\left[\hat{\textbf{a}}\atop \hat{\textbf{b}}\right]}\, \vartheta_{\psi^\mu}\left[a\atop b\right]\,\vartheta_{\chi^{1,2}}\left[a+h_1\atop b+g_1\right]\,\vartheta_{\chi^{3,4}}\left[a+h_2\atop b+g_2\right]\,\vartheta_{\chi^{5,6}}\left[a-h_1-h_2\atop b-g_1-g_2\right]  \\
			&\times \left|\vartheta_{\omega^{1}}\left[\gamma_1\atop \delta_1\right]\vartheta_{y^{1}}\left[\gamma_1+h_1\atop \delta_1+g_1\right] \right| \times \left| \vartheta_{\omega^{2}}\left[\gamma_2\atop \delta_2\right]\vartheta_{y^2}\left[\gamma_2+h_1\atop \delta_2+g_1\right]\right| \\
			&\times \left|\vartheta_{\omega^3}\left[\gamma_3\atop \delta_3\right]\vartheta_{y^3}\left[\gamma_3+h_2\atop \delta_3+g_2\right] \right| \times \left| \vartheta_{\omega^4}\left[\gamma_4\atop \delta_4\right]\vartheta_{y^4}\left[\gamma_4+h_2\atop \delta_4+g_2\right]\right| \\
			&\times \left|\vartheta_{\omega^5}\left[\gamma_5\atop \delta_5\right]\vartheta_{y^5}\left[\gamma_5-h_1-h_2\atop \delta_5-g_1-g_2\right] \right| \times \left|\vartheta_{\omega^6}\left[\gamma_6\atop \delta_6\right]\vartheta_{y^6}\left[\gamma_6-h_1-h_2\atop \delta_6-g_1-g_2\right]\right| \\
			 &\times \bar\vartheta_{\bar\psi^{1\ldots 5}}\left[k\atop \ell\right]^5\,\bar\vartheta_{\bar\eta^1}\left[k+h_1\atop \ell+g_1\right]\,\bar\vartheta_{\bar\eta^2}\left[k+h_2\atop \ell+g_2\right]\,\bar\vartheta_{\bar\eta^3}\left[k-h_1-h_2\atop \ell-g_1-g_2\right] 
			 \times\bar\vartheta_{\bar\phi^{1\ldots 4}}\left[\rho\atop\sigma\right]^4\,\bar\vartheta_{\bar\phi^{5\ldots 8}}\left[\rho+H\atop\sigma+G\right]^4\,,
	\end{split}
	\label{IntermediatePFmodel1}
\end{equation}
where we have added a subscript to each theta function to indicate the corresponding free fermions. We now explain the various contributions in this expression. The first line includes the summation over the $\mathbb Z_2$ values of the independent components $A_i$ and $B_i$, the phase $\Phi$ to be fixed later on, and the RNS fermions in $S$, which are suitably twisted by $(h_1,g_1)$ and $(h_2,g_2)$. The second through fourth lines correspond to the auxiliary fermions, while the last line is the contribution of the (complex) Kac-Moody fermions. In this expression, the theta functions are arranged in the same order as the fermions in $\mathds{1}$ (complexified whenever possible), with the exception of the auxiliary (lattice) fermions, which are slightly rearranged for convenience. Absolute values of the form $|\vartheta_\omega\left[\gamma\atop\delta\right]|$ stand for the combined left- and right-contributions $\vartheta_\omega\left[\gamma\atop\delta\right]^{1/2}  \bar\vartheta_{\bar\omega}\left[\gamma\atop\delta\right]^{1/2}$ associated to $\omega$ and $\bar\omega$ respectively, and similarly for the $y,\bar y$ pairs.

At this stage, we may also define the column vectors $\hat{\textbf{a}}, \hat{\textbf{b}}$, which contain the characteristics of the theta functions entering the partition function \eqref{IntermediatePFmodel1}
\begin{equation}
	\begin{split}
		\hat{\textbf{a}} &= (a,\ldots,a-h_1-h_2, \gamma_1,\ldots, \gamma_6-h_1-h_2, k,\ldots, k+h_1, \ldots,k-h_1-h_2,\rho,\ldots,\rho+H,\ldots)  \,,\\
		\hat{\textbf{b}} &= (b,\ldots,b-g_1-g_2, \delta_1,\ldots, \delta_6-g_1-g_2, \ell,\ldots, \ell+g_1, \ldots,\ell-g_1-g_2,\sigma,\ldots,\sigma+G,\ldots)  \,.
	\end{split}
	\label{ExampleThetaArgs}
\end{equation}
As we have seen in the previous section, the phase $\Phi$ can be split into a modular invariant part $\Omega$, and a part $\Lambda$ with a non-trivial transformation law dictated by the dot products in \eqref{ModularityLambda}. Using \eqref{ExampleThetaArgs}, one finds
\begin{equation}
	\begin{split}
		\frac{1}{4}\,\hat{\textbf{a}}\cdot \hat{\textbf{a}}  &= a^2-H^2-2(k^2+\rho^2+H\rho) \to a+H ~~(\text{mod }2)\,,\\
		\frac{1}{2}\,\hat{\textbf{a}}\cdot \hat{\textbf{b}}  &= 2(ab-HG-2k\ell-\rho G-H\sigma-2\rho\sigma) \to 0 ~~(\text{mod }2)\,,
	\end{split}
\end{equation}
which means that the modularity conditions \eqref{ModularityLambda} can be solved by the choice
\begin{equation}
	\Lambda\left[ \hat{\textbf{a}} \atop \hat{\textbf{b}} \right]  = a+b+HG \,.
	\label{Lsolve}
\end{equation}
All remaining freedom is absorbed into the modular invariant phase $\Omega$ and, hence, into the choice of a anti-symmetric matrix $\textbf{M}$ as in \eqref{OmegaSolution}.
To complete the first stage of the map and uniquely fix $\Omega$, we shall next determine the matrix $\textbf{R}$ responsible for change of basis from $\mathscr{B}$ to $\mathscr{B}'$.
This can be easily read off from \eqref{ModelExpandBprime} by reorganising $\alpha$ and $\beta$ as an expansion in terms of the original basis, and identifying the coefficient of the $v_i A_j$-term as $\tilde{R}_{ij}$. Furthermore, we calculate the offset matrix $\textbf{Q}$, by carefully keeping track of the periodicities in the reduction vectors entering \eqref{OffsetPhase}. The result reads
\begin{equation}
	\textbf{R} = \left(
\begin{array}{cccccccccccc}
 1 & 1 & 1 & 1 & 1 & 1 & 1 & 1 & 1 & 0 & 0 & 0 \\
 1 & 0 & 0 & 0 & 0 & 0 & 0 & 0 & 0 & 0 & 0 & 0 \\
 0 & 1 & 0 & 0 & 0 & 0 & 0 & 0 & 0 & 0 & 0 & 0 \\
 0 & 0 & 1 & 0 & 0 & 0 & 0 & 0 & 0 & 0 & 0 & 0 \\
 0 & 0 & 0 & 1 & 0 & 0 & 0 & 0 & 0 & 0 & 0 & 0 \\
 0 & 0 & 0 & 0 & 1 & 0 & 0 & 0 & 0 & 0 & 0 & 0 \\
 0 & 0 & 0 & 0 & 0 & 1 & 0 & 0 & 0 & 0 & 0 & 0 \\
 0 & 0 & 0 & 0 & 0 & 0 & 1 & 0 & 0 & 0 & 0 & 0 \\
 0 & 0 & 0 & 0 & 0 & 0 & 0 & 1 & 0 & 0 & 1 & 0 \\
 0 & 0 & 0 & 0 & 0 & 0 & 0 & 1 & 0 & 1 & 0 & 0 \\
 0 & 0 & 0 & 0 & 0 & 0 & 0 & 0 & 1 & 0 & 0 & 1 \\
 0 & 0 & 0 & 0 & 0 & 0 & 0 & 0 & 0 & 0 & 0 & 1 \\
\end{array}
\right) \, ,\  \textbf{Q}=\left(
\begin{array}{cccccccccccc}
 0 & 0 & 1 & 1 & 1 & 1 & 1 & 1 & 1 & 1 & 1 & 1 \\
 0 & 0 & 1 & 1 & 1 & 1 & 1 & 1 & 1 & 1 & 1 & 1 \\
 1 & 1 & 0 & 0 & 0 & 0 & 0 & 0 & 0 & 0 & 0 & 0 \\
 1 & 1 & 0 & 0 & 0 & 0 & 0 & 0 & 0 & 0 & 0 & 0 \\
 1 & 1 & 0 & 0 & 0 & 0 & 0 & 0 & 0 & 0 & 0 & 0 \\
 1 & 1 & 0 & 0 & 0 & 0 & 0 & 0 & 0 & 0 & 0 & 0 \\
 1 & 1 & 0 & 0 & 0 & 0 & 0 & 0 & 0 & 0 & 0 & 0 \\
 1 & 1 & 0 & 0 & 0 & 0 & 0 & 0 & 0 & 0 & 0 & 0 \\
 1 & 0 & 0 & 0 & 0 & 0 & 0 & 0 & 1 & 0 & 0 & 0 \\
 1 & 0 & 0 & 0 & 0 & 0 & 0 & 0 & 0 & 1 & 0 & 0 \\
 1 & 1 & 0 & 0 & 0 & 0 & 0 & 0 & 0 & 0 & 1 & 1 \\
 1 & 1 & 0 & 0 & 0 & 0 & 0 & 0 & 0 & 0 & 1 & 1 \\
\end{array}
\right) \,.
	\label{RQmatrices}
\end{equation}

So far, the analysis has relied only on the choice of basis. For the purposes of illustration, we took the symmetric basis \eqref{SymmetricBasis} as our starting point, and systematically worked out the exact map of the one-loop partition function of every free fermionic model in this class, and brought it to the form \eqref{IntermediatePFmodel1}. Indeed, the knowledge of \eqref{RQmatrices} allows the determination of the matrix $\textbf{M}$ from \eqref{Msolution} and, therefore, of the full phase $\Phi$, for any choice of GGSO coefficients defining a consistent free fermionic vacuum. It should be stressed that the phase $\Phi$, determined by this procedure, is unique (modulo 2). Moreover, at this stage, the map works both ways: any choice of the anti-symmetric matrix $\mathbf{M}$ satisfying \eqref{Mcond} gives rise to a consistent free fermionic vacuum corresponding to the symmetric basis.

We are not yet done. The form \eqref{IntermediatePFmodel1} is ``intermediate", in the sense that the contribution of the auxiliary fermions in  has not yet been cast into twisted/shifted Narain lattice form. 
We now proceed to work this out in some detail. For simplicity, we focus on the contribution of the first twisted plane, namely the auxiliary fermions $y^1,\omega^1$ and their right-moving counterparts, since the analysis for the remaining auxiliary fermions is completely analogous. To this end, we first define the block
\begin{equation}
	\hat\Gamma\left[\begin{array}{c|c}H_1 & h_1 \\ G_1 & g_1\end{array}\right] = \frac{1}{2}\sum_{\gamma_1,\delta_1=0,1}  \left|\vartheta_{\omega^{1}}\left[\gamma_1\atop \delta_1\right]\vartheta_{y^{1}}\left[\gamma_1+h_1\atop \delta_1+g_1\right] \right| \,(-1)^{\gamma_1 G_1+\delta_1 H_1+H_1 G_1} \,,
	\label{LatticeBlock}
\end{equation}
that includes the theta function contributions of the aforementioned fermions to the partition function. In what follows, we will first check that \eqref{LatticeBlock} is identified with a (1,1) lattice twisted by the $\mathbb Z_2$ associated to $(h_1,g_1)$, and shifted by an additional freely-acting $\mathbb Z_2$ associated to $(H_1,G_1)$ which shifts along the circle parametrised by $X^1$. Subsequently, we will see how the auxiliary ``lattice'' fermion contributions to the partition function can be systematically organised in terms of orbifold blocks of this type.

Consider first the block \eqref{LatticeBlock} in the sectors $(h_1,g_1)\neq (0,0)$, which involves twisted boundary conditions for $y^1, \bar y^1$, but not for $\omega^1, \bar\omega^1$. As discussed in Section \eqref{MExpample}, this implies twisted boundary conditions for the corresponding circle coordinate $X^1$. Making use of the triple product identity
\begin{equation}
	\left| \vartheta\left[\gamma_1 \atop \delta_1\right] \vartheta\left[\gamma_1+h_1 \atop \delta_1+g_1\right] \vartheta\left[1-h_1 \atop 1-g_1\right] \right| = \left| 2\,\eta^3 \right|\,,
\end{equation}
valid whenever the l.h.s. is non-vanishing, \emph{i.e.} for $(\gamma_1,\delta_1)\neq(1,1)$ and $(\gamma_1+h_1,\delta_1+g_1)\neq(1,1)$ (mod 2), the block \eqref{LatticeBlock} becomes
\begin{equation}
	\hat\Gamma\left[\begin{array}{c|c}H_1 & h_1 \\ G_1 & g_1\end{array}\right] = \left|\frac{2\,\eta^3}{\vartheta\left[1-h_1\atop 1-g_1\right]}\right|   \sum_{(\gamma_1,\delta_1)\neq (1,1)\atop (\gamma_1+h_1,\delta_1+g_1)\neq(1,1)} \frac{1}{2}\,(-1)^{(\gamma_1+H_1)(\delta_1+ G_1)+\gamma_1\delta_1} \,.
\end{equation}
Now notice that if $(H_1,G_1)=(0,0)$ or $(H_1,G_1)=(h_1,g_1)$ (mod 2), the phase inside the sum trivialises as a result of the two independent constraints on the summation, and one obtains simply\footnote{The appearance of the cubic power of the Dedekind function is an artefact of our convention of factoring out all left- and right-moving eta functions in the partition function \eqref{IntermediatePFmodel1}.} $\left|2\,\eta^3 /\vartheta\left[1-h_1\atop 1-g_1\right]\right|$, where the factor of 2 correctly counts the two fixed points of the first circle under $X^1\to -X^1$. For all other choices of $(H_1,G_1)$ instead, it is easy to see that $\hat\Gamma$ vanishes. Taken together, these properties imply that $\hat\Gamma$ precisely matches the contribution to the partition function of a compact boson $X^1$, twisted by a rotational $\mathbb Z_2$ orbifold parametrised by $(h_1,g_1)$, and shifted by a freely-acting $\mathbb Z_2$ orbifold parametrised by $(H_1,G_1)$. In other words, $\hat\Gamma$ correctly reduces to the twisted/shifted lattice of the corresponding orbifold theory in sectors $(h_1,g_1)\neq (0,0)$. 

To complete the identification, we must also consider the case $(h_1,g_1)=(0,0)$ where the twist is absent. In this case, $\hat\Gamma$ contains the product $\vartheta \bar\vartheta$, which can be ``bosonised" using the sum representation of theta functions. Specifically, denoting by $m_1, n_1$ the integers entering the sums of the left- and right-handed theta functions, respectively, we have
\begin{equation}
		\hat\Gamma\left[\begin{array}{c|c}H_1 & 0 \\ G_1 & 0\end{array}\right] = \frac{1}{2}\sum_{\gamma_1,\delta_1=0,1} (-1)^{\gamma_1 G_1+\delta_1 H_1+H_1 G_1} \sum_{m_1,n_1\in\mathbb Z}  (-1)^{\delta_1 n_1} \,q^{\alpha' P_L^2/4}\,\bar q^{\alpha'P_R^2/4} \,,
\end{equation}
with
\begin{equation}
	P_L^2 = \left(\frac{m_1-\gamma_1/2-n_1/2}{\sqrt{\frac{\alpha'}{2}}} + \sqrt{\frac{\alpha'}{2}}\,\frac{n_1}{\alpha'}\right)^2 \ ,\quad P_R^2 = \left(\frac{m_1-\gamma_1/2-n_1/2}{\sqrt{\frac{\alpha'}{2}}} - \sqrt{\frac{\alpha'}{2}}\,\frac{n_1}{\alpha'}\right)^2 \,,
\end{equation}
and we have also shifted the summation variable $n_1\to m_1-n_1$. This looks very similar to the standard form of a shifted Narain lattice at the fermionic radius $R=\sqrt{\alpha'/2}$, except for the presence of $\gamma_1, \delta_1$, which we still need to get rid of. This is accomplished as follows. First, define $m_1' = 2m_1-\gamma_1$ and notice that we can rewrite the phase $(-1)^{\gamma_1 G_1}$ as $(-1)^{m_1' G_1}$, while $\gamma_1$ is completely absorbed into $m_1'$ in the lattice momenta $P_L, P_R$. The joint summation over $\gamma_1=0,1$ and $m_1\in\mathbb Z$ can then be traded for an independent summation over $m_1' \in \mathbb Z$. Furthermore, the summation over $\delta_1=0,1$ acts as a projector, enforcing the constraint $n_1-H_1=0$ (mod 2). We solve this by setting $n_1-H_1 = 2n_1'$ in terms of an independent integer $n_1'\in\mathbb Z$, and we obtain
\begin{equation}
		\hat\Gamma\left[\begin{array}{c|c}H_1 & 0 \\ G_1 & 0\end{array}\right] = \sum_{m_1',n_1'\in\mathbb Z}  (-1)^{(m_1'+H_1) G_1} \,q^{\frac{\alpha'}{4}\left(\frac{m_1'-H_1-2n_1'}{\sqrt{2\alpha'}} + \sqrt{2\alpha'}\,\frac{n_1'+\frac{H_1}{2}}{\alpha'}\right)^2}\,\bar q^{\frac{\alpha'}{4}\left(\frac{m_1'-H_1-2n_1'}{\sqrt{2\alpha'}} -\sqrt{2\alpha'}\,\frac{n_1'+\frac{H_1}{2}}{\alpha'}\right)^2} \,.
\end{equation}
Finally, by shifting $m_1' \to m_1'+2n_1'+H_1$, we recover the standard form of the shifted Narain lattice of signature (1,1), corresponding to the momentum shift
\begin{equation}
	X^1_L \to X^1_L + \pi \tilde{R}/2 \ ,\quad X^1_R \to X^1_R + \pi \tilde{R}/2 \,,
	\label{MomentumShiftOrb}
\end{equation}
at the T-dual fermionic radius $\tilde{R} \equiv \alpha'/R= \sqrt{2\alpha'}$. Alternatively, by exchanging the roles of momenta and windings, we can think of this as the winding shift 
\begin{equation}
	X^1_L \to X^1_L + \alpha' \pi /2R \ ,\quad X^1_R \to X^1_R - \alpha'\pi/2R \,,
	\label{WindingShiftOrb}
\end{equation}
at the fermionic radius $R=\sqrt{\alpha'/2}$. Of course, from the point of view of the partition function, both descriptions look completely identical at the fermionic point.

Since the rotational orbifold twists pairs of (real) internal coordinates, it is convenient to join the corresponding $(1,1)$ the lattices together into $(2,2)$ twisted/shifted lattices of the form
\begin{equation}
	\Gamma_{2,2}\left[\begin{array}{cc|c}H_1 & H_2 & h_1 \\ G_1 & G_2 & g_1\end{array}\right] = \frac{1}{2^2}\sum_{\boldsymbol{\gamma},\boldsymbol{\delta}}  \left|\vartheta\left[\gamma_1\atop \delta_1\right]\vartheta\left[\gamma_1+h_1\atop \delta_1+g_1\right]\vartheta\left[\gamma_2\atop \delta_2\right]\vartheta\left[\gamma_2+h_1\atop \delta_2+g_1\right] \right| \,(-1)^{\boldsymbol{\gamma}\cdot\textbf{G}+\boldsymbol{\delta}\cdot\textbf{H}+\textbf{H}\cdot\textbf{G}} \,,
	\label{Fermionic22Lattice}
\end{equation}
where the dot products in the above expression simply run over the corresponding $T^2$ directions, $I=1,2$. The twisted sector $(h_1,g_1)\neq(0,0)$ is simply a product of the corresponding twisted $(1,1)$ ``lattices" mentioned previously, with the corresponding constraints. In the fully untwisted sector $(h_1,g_1)=(0,0)$, one recovers the shifted Narain lattice of signature $(2,2)$
\begin{equation}
	\Gamma_{2,2}\left[\begin{array}{cc|c}H_1 & H_2 & 0 \\ G_1 & G_2 & 0\end{array}\right](T,U) = \sum_{m_1, m_2, n_1, n_2\in \mathbb Z} (-1)^{m_1 G_1+m_2 G_2}\,q^{\alpha' |P_L|^2/4}\,\bar q^{\alpha' |P_R|^2/4} \,,
	\label{ShiftedLatticeDef}
\end{equation}
with complex left-moving lattice momentum
\begin{equation}
	P_L = \frac{m_2-U m_1+\frac{1}{\alpha'}\,T\left(n_1+\frac{H_1}{2}+U(n_2+\frac{H_2}{2})\right)}{\sqrt{T_2 U_2}} \,,
\end{equation}
and the right-moving complex momentum $P_R$ is obtained from $P_L$ by replacing $T\to \bar T$. Here, $T$ is the K\"ahler modulus and $U$ is the complex structure modulus of the target space 2-torus parametrised by $X^1, X^2$. Matching with the fermionic description \eqref{Fermionic22Lattice} then requires\footnote{This value reflects our choice to view the freely-acting orbifolds acting as momentum shifts of the form \eqref{MomentumShiftOrb} along both directions  $X^1, X^2$ of $T^2$.} $T=2i\alpha'$ and $U=i$, which is consistent with the values of a square 2-torus without $B$-field, factorised into a product of two circles at the dual fermionic radius.

After these considerations, all that is left to do is to replace the $(\gamma_I, \delta_I)$ contributions in the partition function (including their thetas), with three products of the (2,2) twisted/shifted lattices.
To this end, we will now ``invert" \eqref{Fermionic22Lattice} by a discrete Fourier transform, and cast it into a suitable form that allows the aforementioned replacement. Multiplying \eqref{Fermionic22Lattice} by $(-1)^{\textbf{G}\cdot\boldsymbol{\epsilon}+\textbf{H}\cdot\boldsymbol{\zeta}+\textbf{H}\cdot\textbf{G}}$, and summing over $H_I,G_I=0,1$ (times the appropriate projector factor $1/4$), turns the sum over $\gamma_I,\delta_I$ into projectors that fix $\gamma_I = \epsilon_I$ and $\delta_I = \zeta_I$, 
\begin{equation}
	\left|\vartheta\left[\epsilon_1\atop \zeta_1\right]\vartheta\left[\epsilon_1+h_1\atop \zeta_1+g_1\right]\vartheta\left[\epsilon_2\atop \zeta_2\right]\vartheta\left[\epsilon_2+h_1\atop \zeta_2+g_1\right] \right| = \frac{1}{4}\sum_{\textbf{H},\textbf{G}} \Gamma_{2,2}\left[\begin{array}{cc|c}H_1 & H_2 & h_1 \\ G_1 & G_2 & g_1\end{array}\right]\,(-1)^{\textbf{G}\cdot\boldsymbol{\epsilon}+\textbf{H}\cdot\boldsymbol{\zeta}+\textbf{H}\cdot\textbf{G}} \,.
	\label{InversionIntermediate}
\end{equation}
This relation is valid for arbitrary integers $\epsilon_I, \zeta_I$ (defined modulo 2). We can, therefore, apply it for $\epsilon_I, \zeta_I \to \gamma_I, \delta_I$, and extend the analysis to the entire six-dimensional lattice, \emph{i.e.} henceforth $I=1,\ldots,6$.
Let us collectively denote by $\|\vartheta \|$ the product of all theta functions in absolute values in \eqref{IntermediatePFmodel1}, corresponding to all auxiliary fermions. Their characteristics involve $\gamma_I, \delta_I$ as well as the twist parameters $h_1,h_2,g_2,g_2$. Applying the inversion formula \eqref{InversionIntermediate} three times, we have
\begin{equation}
	\|\vartheta \| \left[\begin{array}{c|c}\boldsymbol{\gamma} & \textbf{h} \\ \boldsymbol{\delta} & \textbf{g}\end{array}\right]= \frac{1}{2^6} \sum_{\textbf{H},\textbf{G}} \Gamma_{6,6}\left[\begin{array}{c|c}\textbf{H} & \textbf{h} \\ \textbf{G} & \textbf{g}\end{array}\right]\,(-1)^{\textbf{G}\cdot\boldsymbol{\gamma}+\textbf{H}\cdot\boldsymbol{\delta}+\textbf{H}\cdot\textbf{G}} \,,
	\label{InversionIntermediateFull}
\end{equation}
with the dot products in the phase now running over all $I=1,\ldots 6$. For compactness, we used the obvious definition of the (6,6) lattice
\begin{equation}
	\Gamma_{6,6}\left[\begin{array}{c|c}\textbf{H} & \textbf{h} \\ \textbf{G} & \textbf{g}\end{array}\right] = \Gamma_{2,2}\left[\begin{array}{cc|c}H_1 & H_2 & h_1 \\ G_1 & G_2 & g_1\end{array}\right] \Gamma_{2,2}\left[\begin{array}{cc|c}H_3 & H_4 & h_2 \\ G_3 & G_4 & g_2\end{array}\right] \Gamma_{2,2}\left[\begin{array}{cc|c}H_5 & H_6 & h_1+h_2 \\ G_5 & G_6 & g_1+g_2\end{array}\right] \,,
\end{equation}
factorised as a product of three (2,2) twisted/shifted lattices, and with the moduli lying at the fermionic point values for all three 2-torii. 

It is now clear what we need to do. We should first reconstruct the part of the phase $\Phi$ participating in the summation over $\gamma_I, \delta_I$,  include it into \eqref{InversionIntermediateFull}, and then carry out this summation, so that we are only left with the twisted/shifted (6,6) lattice, now summed over $H_I, G_I$. This ``transmutation" procedure by which the original fermionic boundary conditions $\gamma_I,\delta_I$ of auxiliary fermions (and their associated thetas) are turned into standard orbifold lattice form that includes the freely acting parameters $H_I,G_I$, will typically involve a new phase that we have yet to determine. 

For this purpose, we ``dimensionally reduce" the 12-dimensional vectors $A_i, B_i$ of \eqref{BprimeBasisParams}, by splitting the indices $i\to (I,\hat\alpha)$ and $j\to(J,\hat\beta)$, such that the upper-case Latin indices $I,J,\ldots$ run over the lattice directions, while hatted lower-case Greek indices $\hat\alpha, \hat\beta,\ldots$ run over all other directions. For our symmetric orbifold, the only place where $\gamma_I,\delta_I$ enter the phase $\Phi$ is in the modular invariant bilinear $\Omega$, which we decompose as
\begin{equation}
	\Omega = A_i \,M_{ij}\, B_j = \gamma_I \,M_{IJ}\,\delta_J + \gamma_I\, M_{I\hat\beta}\,B_{\hat\beta} + A_{\hat\alpha} \,M_{\hat\alpha J}\,\delta_J + A_{\hat\alpha}\,M_{\hat\alpha \hat\beta}\,B_{\hat\beta} \,.
	\label{OmegaDecompose}
\end{equation}
Notice the slight abuse of notation $A_I \to \gamma_I$ and $B_I\to \delta_I$.
Clearly, only the first three terms participate in the transmutation. Let us denote $X_I \equiv M_{I\hat\alpha} A_{\hat\alpha}$, $Y_I \equiv M_{I\hat\alpha} B_{\hat\alpha}$ and construct the $6\times 6$ symmetric matrix $\Sigma_{IJ}$ (valued in $\mathbb Z_2$) whose upper triangular elements\footnote{In our symmetric orbifold, $\Sigma_{IJ}=\tilde{C}_{IJ}$ (mod 2), since the $\gamma_I,\delta_I$ do not contribute to $\textbf{Q}$, while the corresponding change of basis in \eqref{Msolution} is trivial.} are defined by $\Sigma_{IJ} = M_{IJ}$ (for $I<J$). Furthermore, it is always possible to choose the diagonal elements (zeros or ones) such that $\det(\boldsymbol{\Sigma}) = 1$ (mod 2). With such a choice\footnote{Although not unique, the particular choice of diagonal elements of $\Sigma_{IJ}$ will not affect our results. The only thing that matters is that such a choice can always be made, such that $\Sigma_{IJ}$ is invertible over $\mathbb Z_2$.}, the matrix $\boldsymbol{\Sigma}$ is invertible over the field $\mathbb{Z}_2$, and we denote its inverse by $\boldsymbol{\Sigma}^{-1}$ (which is obviously also $\mathbb Z_2$ valued).

We now multiply \eqref{InversionIntermediateFull} by the relevant ($\gamma_I,\delta_I$-dependent) part of the phase entering the partition function, and sum over $\gamma_I,\delta_I$ to obtain
\begin{equation}
	\begin{split}
		\frac{1}{2^6} & \sum_{\boldsymbol{\gamma},\boldsymbol{\delta}}\|\vartheta \|  \left[\begin{array}{c|c}\boldsymbol{\gamma} & \textbf{h} \\ \boldsymbol{\delta} & \textbf{g}\end{array}\right] (-1)^{\boldsymbol{\gamma}^T \textbf{Y}+\boldsymbol{\delta}^T \textbf{X}+\boldsymbol{\gamma}^T \boldsymbol{\Sigma\delta}} \\
		&= \frac{1}{2^6} \sum_{\textbf{H},\textbf{G}} \Gamma_{6,6}\left[\begin{array}{c|c}\textbf{H} & \textbf{h} \\ \textbf{G} & \textbf{g}\end{array}\right]\,(-1)^{(\textbf{X}+\textbf{H})^T \boldsymbol{\Sigma}^{-1} (\textbf{Y}+\textbf{G})+\textbf{H}^T\textbf{G}} \,\left\{ \frac{1}{2^6} \sum_{\boldsymbol{\gamma},\boldsymbol{\delta}}  (-1)^{(\boldsymbol{\gamma} + \boldsymbol{\Sigma}^{-1} \textbf{X}+\boldsymbol{\Sigma}^{-1}\textbf{H})^T \boldsymbol{\Sigma} (\boldsymbol{\delta}+\boldsymbol{\Sigma}^{-1}\textbf{Y}+\boldsymbol{\Sigma}^{-1}\textbf{G})}     \right\} \,.
	\end{split}
\end{equation}
In this expression, we adopted a convenient matrix notation for the phases where all matrix products are with respect to the lattice indices $I,J$. The quantity in the curly brackets on the r.h.s. is clearly independent of $H_I, G_I, X_I, Y_I$, as one may immediately verify by appropriately shifting the summations of $\gamma_I, \delta_I$. Moreover, carrying out the sum over $\delta_I$ on the r.h.s. yields the product of six projectors setting $\gamma_I' = \Sigma_{IJ}\gamma_J$ to be even. However, since $\Sigma_{IJ}$ is invertible, the six $\gamma_I'$'s are independent and we can equivalently sum over them. The sum over the projectors then is then trivial and equals unity. We therefore obtain the result
\begin{equation}
		\frac{1}{2^6}  \sum_{\boldsymbol{\gamma},\boldsymbol{\delta}}\|\vartheta \|  \left[\begin{array}{c|c}\boldsymbol{\gamma} & \textbf{h} \\ \boldsymbol{\delta} & \textbf{g}\end{array}\right] (-1)^{\boldsymbol{\gamma}^T \textbf{Y}+\boldsymbol{\delta}^T \textbf{X}+\boldsymbol{\gamma}^T \boldsymbol{\Sigma\delta}} = \frac{1}{2^6} \sum_{\textbf{H},\textbf{G}} \Gamma_{6,6}\left[\begin{array}{c|c}\textbf{H} & \textbf{h} \\ \textbf{G} & \textbf{g}\end{array}\right]\,(-1)^{(\textbf{X}+\textbf{H})^T \boldsymbol{\Sigma}^{-1} (\textbf{Y}+\textbf{G}) +\textbf{H}^T \textbf{G}}  \,.
	\label{FermionLatticeInversion}
\end{equation}
One might worry about the freedom to choose the diagonal elements of $\boldsymbol{\Sigma}$ in the above derivation. However, such elements only produce modular invariant phases of the form $(-1)^{\gamma_I\delta_I}$ on the l.h.s., which never contribute to the sum since terms with $\gamma_I=\delta_I=1$ are always accompanied by a vanishing theta function in $\|\vartheta\|$. We will return to this point in Section \ref{LatticeReps}.

We may now apply \eqref{FermionLatticeInversion} to replace the contribution of all auxiliary fermions $y^I,\omega^I, \bar y^I, \bar\omega^I$ in the partition function \eqref{IntermediatePFmodel1} with a product of three twisted/shifted (2,2) lattices, each one associated to its corresponding 2-torus. We then obtain the final expression for the one-loop partition function in orbifold form
\begin{equation}
	\begin{split}
	Z&=\frac{1}{\eta^{12}\,\bar\eta^{24}}\frac{1}{2^{12}}\sum_{\{A'\},\{B'\}} (-1)^{\Phi'\left[ \{A'\}\atop \{ B' \} \right]}\, \vartheta_{\psi^\mu}\left[a\atop b\right]\,\vartheta_{\chi^{1,2}}\left[a+h_1\atop b+g_1\right]\,\vartheta_{\chi^{3,4}}\left[a+h_2\atop b+g_2\right]\,\vartheta_{\chi^{5,6}}\left[a-h_1-h_2\atop b-g_1-g_2\right]  \\
			&\times \Gamma_{2,2}\left[\begin{array}{cc|c}H_1 & H_2 & h_1 \\ G_1 & G_2 & g_1\end{array}\right] \, \Gamma_{2,2}\left[\begin{array}{cc|c}H_3 & H_4 & h_2 \\ G_3 & G_4 & g_2\end{array}\right] \, \Gamma_{2,2}\left[\begin{array}{cc|c}H_5 & H_6 & h_1+h_2 \\ G_5 & G_6 & g_1+g_2\end{array}\right]\\
			 &\times \bar\vartheta_{\bar\psi^{1\ldots 5}}\left[k\atop \ell\right]^5\,\bar\vartheta_{\bar\eta^1}\left[k+h_1\atop \ell+g_1\right]\,\bar\vartheta_{\bar\eta^2}\left[k+h_2\atop \ell+g_2\right]\,\bar\vartheta_{\bar\eta^3}\left[k-h_1-h_2\atop \ell-g_1-g_2\right] 
			 \times\bar\vartheta_{\bar\phi^{1\ldots 4}}\left[\rho\atop\sigma\right]^4\,\bar\vartheta_{\bar\phi^{5\ldots 8}}\left[\rho+H\atop\sigma+G\right]^4\,.
	\end{split}
	\label{OrbifoldPFmodel1}
\end{equation}
Notice that the auxiliary fermion summation parameters $\gamma_I,\delta_I$ have been replaced by the freely-acting orbifold parameters $H_I,G_I$ for all $I=1,\ldots,6$, so that the space of summation variables is now
\begin{equation}
		\{A' \} = \{a,H_1,\ldots,H_6, k,\rho, h_1,h_2, H \} \,,\quad \{B'\} = \{b,H_1,\ldots,H_6, \ell,\sigma, g_1,g_2,G\} \,.
\end{equation}
The new phase $\Phi'$ in \eqref{OrbifoldPFmodel1} is obtained by assembling the contributions of \eqref{Lsolve}, the $\gamma,\delta$-independent part of \eqref{OmegaDecompose}, and the phase arising on the r.h.s. of \eqref{FermionLatticeInversion}, yielding
\begin{equation}
	\Phi' = a+b+HG+A_{\hat\alpha}\,M_{\hat\alpha \hat\beta}\,B_{\hat\beta}+(X_I+H_I)\Sigma^{IJ} (Y_J+G_J)+H_I G_I \,,
	\label{OrbifoldPhaseModel1}
\end{equation}
where $\Sigma^{IJ}$ are the elements of $\boldsymbol{\Sigma}^{-1}$. This completes the map, and brings the free fermionic theory to an orbifold representation. 

For illustration purposes, it is instructive to consider a specific model. To this end, we pick a vacuum with $\mathcal N=1$ supersymmetry, three generations of chiral matter, and gauge group $\text{SO}(10)\times\text{U}(1)^3\times\text{SO}(8)^2$, that was built in \cite{Faraggi:2003yd} using the symmetric basis \eqref{SymmetricBasis} with the following choice of associated GGSO matrix
\begin{equation}
	\tilde{\textbf{C}}=
\left(\begin{array}{c c c c c c c c c c c c}		
     1 &  1 &  1 &  1 &  1 &  1 &  1 &  1 &  1 &  1 &  1 & 1\\
     1 &  1 &  1 &  1 &  1 &  1 &  1 &  1 &  1 &  1 &  1 & 1\\
     1 &  1 &  0 &  0 &  0 &  1 &  1 &  0 &  0 &  0 &  0 & 0\\
     1 &  1 &  0 &  0 &  1 &  0 &  1 &  1 &  0 &  0 &  0 & 0\\
     1 &  1 &  0 &  1 &  0 &  0 &  0 &  0 &  0 &  0 &  0 & 1\\
     1 &  1 &  1 &  0 &  0 &  0 &  0 &  0 &  0 &  0 &  1 & 0\\
     1 &  1 &  1 &  1 &  0 &  0 &  0 &  1 &  0 &  0 &  0 & 1\\
     1 &  1 &  0 &  1 &  0 &  0 &  1 &  0 &  0 &  0 &  1 & 0\\
     1 &  0 &  0 &  0 &  0 &  0 &  0 &  0 &  1 &  0 &  0 & 0\\
     1 &  0 &  0 &  0 &  0 &  0 &  0 &  0 &  0 &  1 &  0 & 0\\
     1 &  1 &  0 &  0 &  0 &  1 &  0 &  1 &  0 &  0 &  1 & 1\\
     1 &  1 &  0 &  0 &  1 &  0 &  1 &  0 &  0 &  0 &  1 & 1\\
\end{array}\right) \ , \ \textbf{M}=\left(
\begin{array}{c c c c c c c c c c c c}
 1 & 0 & 0 & 0 & 0 & 0 & 0 & 0 & 0 & 0 & 0 & 0 \\
 0 & 0 & 0 & 0 & 1 & 1 & 0 & 0 & 0 & 0 & 0 & 0 \\
 0 & 0 & 0 & 1 & 0 & 1 & 1 & 1 & 0 & 1 & 1 & 0 \\
 0 & 0 & 1 & 0 & 0 & 0 & 0 & 0 & 1 & 0 & 0 & 1 \\
 0 & 1 & 0 & 0 & 0 & 0 & 0 & 0 & 1 & 0 & 0 & 0 \\
 0 & 1 & 1 & 0 & 0 & 0 & 1 & 0 & 1 & 0 & 0 & 1 \\
 0 & 0 & 1 & 0 & 0 & 1 & 0 & 1 & 1 & 1 & 1 & 0 \\
 0 & 0 & 1 & 0 & 0 & 0 & 1 & 0 & 0 & 0 & 0 & 0 \\
 0 & 0 & 0 & 1 & 1 & 1 & 1 & 0 & 0 & 0 & 0 & 0 \\
 0 & 0 & 1 & 0 & 0 & 0 & 1 & 0 & 0 & 0 & 0 & 0 \\
 0 & 0 & 1 & 0 & 0 & 0 & 1 & 0 & 0 & 0 & 0 & 0 \\
 0 & 0 & 0 & 1 & 0 & 1 & 0 & 0 & 0 & 0 & 0 & 0 \\
\end{array}
\right) \,.
\label{GGSOmodel}
\end{equation}
Here, we displayed also the corresponding $\textbf{M}$-matrix of the model, which is obtained from the above $\tilde{\textbf{C}}$-matrix with the help of \eqref{RQmatrices} and \eqref{Msolution}. Already the determination of $\textbf{M}$ uniquely fixes the phase $\Phi$ in \eqref{IntermediatePFmodel1}, but we shall not explicitly display it here.  Instead, we proceed to determine $\Phi'$ using \eqref{OrbifoldPhaseModel1}. In this case, $\Sigma_{IJ}$ is invertible
\begin{equation}
	\Sigma_{IJ}=\left(
\begin{array}{cccccc}
 0 & 0 & 0 & 1 & 1 & 0 \\
 0 & 0 & 1 & 0 & 1 & 1 \\
 0 & 1 & 0 & 0 & 0 & 0 \\
 1 & 0 & 0 & 0 & 0 & 0 \\
 1 & 1 & 0 & 0 & 0 & 1 \\
 0 & 1 & 0 & 0 & 1 & 0 \\
\end{array}
\right) \ , \quad \Sigma^{IJ}=\left(
\begin{array}{cccccc}
 0 & 0 & 0 & 1 & 0 & 0 \\
 0 & 0 & 1 & 0 & 0 & 0 \\
 0 & 1 & 0 & 1 & 1 & 1 \\
 1 & 0 & 1 & 0 & 0 & 1 \\
 0 & 0 & 1 & 0 & 0 & 1 \\
 0 & 0 & 1 & 1 & 1 & 0 \\
\end{array}
\right) \,,
\end{equation}
Using \eqref{OrbifoldPhaseModel1}, one extracts the phase $\Phi'$ in the orbifold representation
\begin{equation}
	\begin{split}
		\Phi' =& a+b+HG+[ab]+[\rho\sigma]+[(\rho+H)(\sigma+G)] \\
			+& [kG+\ell H+ HG] + [k G_4+\ell H_4+H_4 G_4] + [kG_5+\ell H_5+H_5 G_5] \\
			+& [\rho G_1+\sigma H_1+H_1 G_1]+[\rho G_2+\sigma H_2+H_2 G_2] \\
			+& [\rho G_3+\sigma H_3+H_3 G_3]+ [\rho G_6+\sigma H_6+H_6 G_6]\\
			+& [(h_1+h_2)(G+G_4+G_5)+ (g_1+g_2)(H+H_4+H_5)]  \\
			+& [H(G_2+G_3+G_4+G_5)+G(H_2+H_3+H_4+H_5)] \\
			+& [1,4]+[2,3]+[3,4]+[3,5]+[3,6]+[4,6]+[5,6] \,,
	\end{split}
\end{equation}
where $[I,J]\equiv H_I G_J+G_I H_J$ are ``discrete torsions" for the lattice shift orbifolds, and the terms inside brackets are separately modular invariant. In addition to the $\Lambda$ contributions, the first line involves chirality convention flips, while the last three lines involve discrete torsion terms. Aside from the familiar $\mathbb{Z}_2\times\mathbb{Z}_2$ orbifold action rotating the first and second orbifold planes, respectively, the action of the shift orbifolds can be read off directly from the phase. For example, the shift orbifolds associated to $(H_I,G_I)$ act as $(-1)^{F_{1,2}}\sigma_I$, with $\sigma_I$ being a momentum shift along the $I$th direction of $T^6$, while $F_1,F_2$ are the ``fermion numbers" associated to the spinorial representation of the first and second $\text{E}_8$, respectively. 

As a last cross-check, we may also compute and compare the contributions $Z_{\text{b}}$, $Z_{\text{f}}$ of bosonic and fermionic states to the partition function in the free-fermionic and orbifold representations (at the fermionic point), as expansions in $q,\bar{q}$. Indeed, one finds
\begin{equation}
	\begin{split}
		Z_{\text{b}}=-Z_{\text{f}} = 2\,\bar{q}^{-1}&+872+120\,q^{1/2}\bar{q}^{-1/2}+ 32\,q\,\bar{q}^{-1} \\ 
				&+4080\, q^{1/8}\,\bar{q}^{1/8} + 16\,q^{9/8}\,\bar{q}^{-7/8}+704\,q^{5/8}\,\bar{q}^{-3/8} + \ldots \,,
	\end{split}
\end{equation}
and the contributions match in both formalisms, as they should.

Although further deformations are possible, such as for instance turning on Wilson lines, they will not be discussed here. One of the questions that one may wish to address with the help of \eqref{OrbifoldPFmodel1} and \eqref{OrbifoldPhaseModel1} is related to spacetime supersymmetry. To this end, it is sufficient to study how the R-symmetry charges $a,b$ couple to other boundary conditions in $\Phi'$. Notice first that couplings of $(a,b)$ to the rotations $(h_i,g_i)$ correspond to chirality redefinitions for the spinor characters arising in the decomposition of the theta functions associated to the RNS fermions, and will hence play no role. Couplings of $(a,b)$ to lattice shift parameters $(H_I,G_I)$ correspond to Scherk-Schwarz supersymmetry breaking, correlating the spacetime fermion parity $(-1)^F$ with freely-acting shifts in the internal space. On the other hand, it is not difficult to see that couplings of $(a,b)$ to $(H,G)$, $(k,\ell)$ or $(\rho,\sigma)$ correspond to an explicit breaking of supersymmetry since, even in the decompactification limit, the corresponding gravitino remains massive. For instance, the gravitino arises from the sector $a=1$ and $k=\rho=H=h_1=h_2=H_I=0$. If we study $\Phi'$ in this sector, the various arguments of lower characteristics indicate chirality or other projections. In particular, any coupling of the type $aG_I$ entering $\Phi'$ in this sector, upon summation of $G_I$, will project onto states with odd momentum number $m_I$. Given that the gravitino must be level matched, $|P_L|=|P_R|$, we must also pick vanishing windings for this state. The lowest state associated to the gravitino can then be identified and its mass can be expressed as a function of the compactification moduli. We refer the reader to \cite{Florakis:2016ani} for a further discussion on the gravitino mass and the T-duality properties relevant for free-fermionic models in this class.


\section{Shifted lattices, phases and equivalent representations}
\label{LatticeReps}

In the previous section, we encountered an apparent ambiguity in the definition of $\Sigma_{IJ}$. Specifically, we mentioned that we were free to choose the diagonal elements $\Sigma_{II}$, such that the matrix became invertible in the field $\mathbb Z_2$, \emph{i.e.} such that $\det(\boldsymbol{\Sigma})=1$ (mod 2). In general, this choice is not unique and typically leads to different inverse elements $\Sigma^{IJ}$ which, in turn, lead to different resulting phases $\Phi'$ via \eqref{OrbifoldPhaseModel1}. Therefore, it might appear that the phase $\Phi'$ in the orbifold representation is not uniquely fixed (even though $\Phi$ is) and one might wonder whether distinct choices of invertible matrices $\Sigma_{IJ}$ lead to distinct orbifold theories. It turns out that there exist different equivalent representations of the orbifold shifts, corresponding to the freedom of performing the sums over $\gamma_I$ and $\delta_I$ in \eqref{FermionLatticeInversion} (with or without the diagonal $\gamma_I\delta_I$ terms in the corresponding phases), to the freedom of choosing equivalent lattice vectors, and to the T-duality symmetries of the shifted lattice.

To illustrate this equivalence, we shall consider a class of toy models in 8 non-compact dimensions, arising from the symmetric basis \eqref{SymmetricBasis} after removing the rotational twists $v_9, v_{10}$ and the basis vectors $v_5,\ldots,v_8$ associated to the internal fermions $y^I,\omega^I$ with $I=3,\ldots,6$. This will involve a two-dimensional lattice with $\mathbb Z_2$ momentum shifts, and will allow us to highlight how different representations of the shifts may correspond to different phase insertions in the partition function. The partition function in its intermediate form reads
\begin{equation}
	\begin{split}
	Z=\frac{1}{\eta^{12}\,\bar\eta^{24}}\,\frac{1}{2^{6}} \sum_{\{A\},\{B\}} &(-1)^{a+b+HG+\Omega}\, \vartheta_{\psi^\mu,\chi^I}\left[a\atop b\right]^4  \,\left|\vartheta_{y^1,\omega^{1}}\left[\gamma_1\atop \delta_1\right]^2 \vartheta_{y^2,\omega^{2}}\left[\gamma_2\atop \delta_2\right]^2 \right| \\
			 &\times \bar\vartheta_{\bar\psi^{1\ldots 5},\bar\eta^{1,2,3}}\left[k\atop \ell\right]^8 
			 \,\bar\vartheta_{\bar\phi^{1\ldots 4}}\left[\rho\atop\sigma\right]^4\,\bar\vartheta_{\bar\phi^{5\ldots 8}}\left[\rho+H\atop\sigma+G\right]^4\,.
	\end{split}
	\label{IntermediatePFmodel2}
\end{equation}
Here, we define the 6-dimensional column vectors $\textbf{A}=(a,k,\rho,H,\gamma_1,\gamma_2)^T$ and $\textbf{B}=(b,\ell,\sigma,G,\delta_1,\delta_2)^T$, and the phase is fixed in terms of $\Omega = \textbf{A}^T \textbf{M} \textbf{B}$. The models are then defined by the $6\times 6$ symmetric matrix $\textbf{M}$ containing a priori 21 $\mathbb Z_2$-valued elements.  There are 5 independent modularity conditions arising from $\tau\to\tau+1$ symmetry, which can be explicitly solved (modulo 2) as
\begin{equation}
	\begin{split}
		M_{12} &= M_{23}+M_{25}+M_{26} \,,\\
		M_{13} &= M_{23}+M_{35}+M_{36} \,,\\
		M_{14} &= M_{24}+M_{34}+M_{44} +M_{45}+M_{46} \,,\\
		M_{15} &= M_{25}+M_{35}+M_{56} \,,\\
		M_{16} &= M_{26}+M_{36}+M_{56} \,.
	\end{split}
\end{equation}
Therefore, the modular invariant phase may be parametrised in terms of 16 independent matrix elements as
\begin{equation}
	\begin{split}
		\Omega =& \gamma_1\left[ (M_{25}+M_{35}+M_{56})b + M_{25}\ell + M_{35}\sigma + M_{45}G\right]  \\
			     +& \delta_1 \left[ (M_{25}+M_{35}+M_{56})a + M_{25}k + M_{35}\rho+ M_{45}H \right] \\
			     +& \gamma_2 \left[ (M_{26}+M_{36}+M_{56})b + M_{26}\ell + M_{36}\sigma+ M_{46}G \right]\\
			     +& \delta_2 \left[ (M_{26}+M_{36}+M_{56})a + M_{26}k+M_{36}\rho + M_{46}H\right] \\
			     +& M_{56}(\gamma_1\delta_2+\delta_1\gamma_2) + M_{55}\gamma_1\delta_1 + M_{66}\gamma_2\delta_2  + \ldots \,,
	\end{split}
\end{equation}	
where the ellipses denote $\gamma,\delta$-independent terms that are irrelevant for our discussion. For simplicity, we can define
\begin{equation}
	\begin{split}
		&X_1 = (M_{25}+M_{35}+M_{56})b + M_{25}\ell + M_{35}\sigma + M_{45}G \,,\\
		& X_2 = (M_{26}+M_{36}+M_{56})b + M_{26}\ell + M_{36}\sigma+ M_{46}G \,,\\
		&Y_1 = (M_{25}+M_{35}+M_{56})a + M_{25}k + M_{35}\rho+ M_{45}H \,,\\
		&Y_2 = (M_{26}+M_{36}+M_{56})a + M_{26}k+M_{36}\rho + M_{46}H \,,
	\end{split}
\end{equation}
so that $\Omega$ can be written as
\begin{equation}
	\Omega = \boldsymbol{\gamma}\cdot \textbf{Y}+ \boldsymbol{\delta}\cdot\textbf{X} + M_{56}(\gamma_1\delta_2+\delta_1\gamma_2) + M_{55}\gamma_1\delta_1 + M_{66}\gamma_2\delta_2  + \ldots \,.
\end{equation}
In what follows, we shall consider and treat various choices for $M_{56}, M_{55}, M_{66}, \textbf{X}, \textbf{Y}$ and see how the different ways of summing the fermionic boundary conditions $\gamma_I,\delta_I$ lead to the same shifted orbifold theories, despite the different phase representations.

\paragraph{Case I, the trivial (unshifted) lattice:}

Consider first the trivial case where there is no coupling of $\gamma,\delta$ in $\Omega$, \emph{i.e.} $M_{56}=M_{55}=M_{66}=M_{25}=M_{35}=M_{45}=M_{26}=M_{36}=M_{46}=0$, implying also $X_I=Y_I=0$. Using the two-dimensional analogue of \eqref{InversionIntermediateFull} and directly summing\footnote{Here we do not insert $\gamma_1\delta_1$ or $\gamma_2\delta_2$ terms.} over $\gamma_{1,2}$ and $\delta_{1,2}$, generates four projectors that set $H_1=H_2=G_1=G_2=0$, and one immediately recovers the (2,2) unshifted Narain lattice $\Gamma_{2,2}\left[0~0\atop 0~0\right](2i\alpha',i)$. The same result can be reached using the matrix $\Sigma_{IJ}$ of Section \ref{SymExample}. A priori, in this case $\Sigma_{IJ}$ would be identically zero, so we need to add the diagonal terms $\gamma_1\delta_1$ and $\gamma_2\delta_2$ into $\Omega$ in order to render $\Sigma_{IJ}$ invertible, \emph{i.e.} $\Sigma_{IJ}=\delta_{IJ}$. This does not affect the l.h.s. of \eqref{FermionLatticeInversion}, because the would-be contributions of the extra phases $(-1)^{\gamma_I\delta_I}$ vanish, since for $\gamma_I=\delta_I=1$ they are multiplied by $\vartheta\left[1\atop 1\right]$'s.
Carrying out the summation over $\gamma_{1,2},\delta_{1,2}$ \`a la \eqref{FermionLatticeInversion} produces no new phase, and we are therefore left to sum
\begin{equation}
	\frac{1}{2^2}\sum_{H_1,H_2\atop G_1,G_2} \Gamma_{2,2}\left[\begin{array}{cc}H_1 & H_2  \\ G_1 & G_2 \end{array}\right](2i\alpha', i) = \Gamma_{2,2}\left[\begin{array}{cc}0 & 0  \\ 0 & 0 \end{array}\right](2i\alpha', i) \,,
\end{equation}
hence, indeed recovering the very same unshifted Narain lattice. The above identity may be easily derived by writing the lattice in the Hamiltonian representation, using the sum over $H_{1,2}$ to effectively turn $n_{1,2}+H_{1,2}/2\to n_{1,2}/2$, performing the projector sum over $G_{1,2}$ to effectively replace $m_{1,2}\to 2m_{1,2}$, and finally using the T-duality (inversion) symmetry on the K\"ahler modulus. 

\paragraph{Case II, no mixing:}
Let us now consider the first non-trivial case $M_{56}=M_{55}=M_{66}=0$, where $\gamma_{1,2}$ and $\delta_{1,2}$ have no self-couplings in the phase $\Omega$, but do appear in the phase through their couplings to (non-vanishing) $X_{1,2}$ and $Y_{1,2}$.
A direct summation\footnote{Again, here we do not insert $\gamma_1\delta_1$ or $\gamma_2\delta_2$ terms in the phase.} over $\gamma_{1,2}$ and $\delta_{1,2}$ along the lines of \eqref{InversionIntermediateFull} produces four projectors which fix $H_{1,2}=X_{1,2}$ and $G_{1,2}=Y_{1,2}$
\begin{equation}
	\frac{1}{2^2}\sum_{\gamma_1,\gamma_2\atop \delta_1,\delta_2}\,\left|\vartheta\left[\gamma_1\atop \delta_1\right]^2 \vartheta\left[\gamma_2\atop \delta_2\right]^2 \right| (-1)^{\gamma_1 Y_1+\gamma_2 Y_2+\delta_1 X_1+\delta_2 X_2} = (-1)^{X_1 Y_1+X_2 Y_2}\, \Gamma_{2,2}\left[\begin{array}{cc}X_1 & X_2  \\ Y_1 & Y_2 \end{array}\right](2i\alpha', i)\,.
\end{equation}
Starting from the explicit expression \eqref{ShiftedLatticeDef} for the momentum shifted lattice, we decompose the sum over $m_{1,2}\to 2m_{1,2}'+H_{1,2}'$ into a sum involving new unconstrained momenta $m_{1,2}' \in\mathbb Z$ and a sum over new $\mathbb Z_2$ variables $H_{1,2}'$. Subsequently, we insert the projectors
\begin{equation}
	\frac{1}{2^2}\sum_{G_1', G_2'} (-1)^{G_1' n_1' +G_2' n_2'} \,,
\end{equation}
inside the lattice sum (the $G_{1,2}'$ now act as discrete Lagrange multipliers), which allows us to replace $n_{1,2}\to n_{1,2}'/2$. Finally, we shift the summation $n_{1,2}' \to n_{1,2}'- X_{1,2}$ to remove the $X$'s from the lattice momenta, and we find
\begin{equation}
	(-1)^{X_1 Y_1+X_2 Y_2}\, \Gamma_{2,2}\left[\begin{array}{cc}X_1 & X_2  \\ Y_1 & Y_2 \end{array}\right](2i\alpha', i) = \frac{1}{2^2}\sum_{H_1', H_2'\atop G_1', G_2'} (-1)^{\textbf{H}'\cdot\textbf{Y}+\textbf{G}'\cdot\textbf{X}+\textbf{X}\cdot\textbf{Y}} \,\Gamma_{2,2}'\left[\begin{array}{cc}H_1' & H_2'  \\ G_1' & G_2' \end{array}\right](i\alpha'/2, i)\,.
\end{equation}
The prime on the (2,2) lattice on the r.h.s. implies that this lattice involves two independent winding shifts of the form \eqref{WindingShiftOrb} along the corresponding lattice directions. Again, invoking the T-duality symmetry in the K\"ahler modulus, we can cast the latter into the standard form of a momentum shifted lattice -- a process which also resets the moduli to their dual fermionic values $T=2i\alpha'$ and $U=i$ (\emph{i.e.} the values we started with). The result then reads
\begin{equation}
	\frac{1}{2^2}\sum_{\gamma_1,\gamma_2\atop \delta_1,\delta_2}\,\left|\vartheta\left[\gamma_1\atop \delta_1\right]^2 \vartheta\left[\gamma_2\atop \delta_2\right]^2 \right| (-1)^{\gamma_1 Y_1+\gamma_2 Y_2+\delta_1 X_1+\delta_2 X_2} = \frac{1}{2^2}\sum_{H_1, H_2\atop G_1, G_2} (-1)^{\textbf{H}\cdot\textbf{Y}+\textbf{G}\cdot\textbf{X}+\textbf{X}\cdot\textbf{Y}} \,\Gamma_{2,2}\left[\begin{array}{cc}H_1 & H_2  \\ G_1 & G_2 \end{array}\right](2i\alpha', i)\,.
\end{equation}
It is straightforward to check that the exact same result can be recovered by simply picking the non-singular matrix $\Sigma_{IJ}=\delta_{IJ}$ and applying the summation formula \eqref{FermionLatticeInversion}.

In both Cases I and II treated so far, the lattice directions were uncoupled, in the sense that the off-diagonal elements mixing the corresponding $\gamma,\delta$ were absent since $M_{56}=0$. As a result, the insertion (or not) of diagonal elements in the $\gamma,\delta$ sums led to seemingly different shifted lattices (with winding versus momentum shifts at a different point in moduli space). However, upon invoking the T-duality symmetry of the shifted lattices, it was easy to see that the results match exactly, as expected. In doing so, there was never any difficulty with the phases. We will see below that, if the lattice directions are instead mixed, one naively obtains different phase representations and some more work is necessary in order to identify their equivalence.

\paragraph{Case III, mixing:}
Finally, we will consider the case of $\gamma,\delta$ self-coupling $M_{56}=1$, while keeping $M_{55}=M_{66}=0$. This corresponds to $\boldsymbol{\Sigma}=\left( 0 ~ 1\atop 1 ~ 0\right)$, which is already invertible and so the summation formula \eqref{FermionLatticeInversion} applies directly. Doing so, one readily obtains
\begin{equation}
	\begin{split}
	&\frac{1}{2^2}\sum_{\gamma_1,\gamma_2\atop \delta_1,\delta_2}\,\left|\vartheta\left[\gamma_1\atop \delta_1\right]^2 \vartheta\left[\gamma_2\atop \delta_2\right]^2 \right| (-1)^{\boldsymbol{\gamma}\cdot \textbf{Y}+ \boldsymbol{\delta}\cdot\textbf{X} + \gamma_1\delta_2+\delta_1\gamma_2} \\
		&= \frac{1}{2^2}\sum_{H_1, H_2\atop G_1, G_2} (-1)^{(G_1 X_2+H_1 Y_2+G_1 H_1)+(G_2 X_1+H_2 Y_1+G_2 H_2)+(G_1 H_2+G_2 H_1)+(Y_1 X_2+X_1 Y_2)} \,\Gamma_{2,2}\left[\begin{array}{cc}H_1 & H_2  \\ G_1 & G_2 \end{array}\right](2i\alpha', i)\,.
		\label{Case3mix1}
	\end{split}
\end{equation}
Let us now see how different ways of treating the summation would lead to seemingly different results, and how the latter can be reconciled. For example, instead of using \eqref{FermionLatticeInversion}, we could apply \eqref{InversionIntermediateFull}
to replace the $\gamma,\delta$-dependent theta functions with a sum over (2,2) lattices, and then deal with the $\gamma,\delta$ summations one pair at a time. In this case, the relevant quantity reads
\begin{equation}
	\frac{1}{2^4}\sum_{H_1,H_2\atop G_1,G_2}\sum_{\gamma_1,\gamma_2\atop \delta_1,\delta_2}(-1)^{\gamma_1(Y_1+G_1+\delta_2)+\delta_1(X_1+H_1+\gamma_2)+ \gamma_2(Y_2+G_2)+\delta_2(X_2+H_2)+H_1 G_1+H_2 G_2} \,\Gamma_{2,2}\left[\begin{array}{cc}H_1 & H_2  \\ G_1 & G_2 \end{array}\right](2i\alpha', i)\,.
\end{equation}
The phase contains no $\gamma_1\delta_1$ term, so that summing over $\gamma_1,\delta_1=0,1$ produces two projectors (absorbing a factor of $1/4$) that set $\gamma_2+X_1+H_1 = \delta_2+Y_1+G_1=0$ (mod 2). One way of implementing this constraint is to solve it for $\gamma_2$ and $\delta_2$, which eliminates the corresponding summations, and then plug it into the rest to recover the r.h.s. of \eqref{Case3mix1} with the exact same phase. However, one could alternatively solve the constraint for $H_1,G_1$ and substitute it in the above, to obtain
\begin{equation}
	\frac{1}{2^2}\sum_{H_2,\gamma_2\atop G_2,\delta_2}(-1)^{ \gamma_2(Y_2+G_2)+\delta_2(X_2+H_2)+(X_1+\gamma_2)(Y_1+\delta_2)+H_2 G_2} \,\Gamma_{2,2}\left[\begin{array}{cc}X_1+\gamma_2 & H_2  \\ Y_1+\delta_2 & G_2 \end{array}\right](2i\alpha', i)\,.
	\label{Case3mix2}
\end{equation}
Now $X_1, Y_1, \gamma_2,\delta_2$ appear explicitly in the arguments of the lattice and we can no longer sum over $\gamma_2,\delta_2$. To this end, one may bring $X_1,Y_1,\gamma_2,\delta_2$ back into the phase, by employing the following identity
\begin{equation}
	\Gamma_{2,2}\left[\begin{array}{cc}X_1+\gamma_2 & H_2  \\ Y_1+\delta_2 & G_2 \end{array}\right](2i\alpha', i) = \frac{1}{2}\sum_{H_1',G_1'} (-1)^{H_1'(Y_1+\delta_2)+G_1'(X_1+\gamma_2)} \,\Gamma_{2,2}\left[\begin{array}{cc}H_1' & H_2  \\ G_1' & G_2 \end{array}\right](2i\alpha', i) \,.
	\label{IdentityLattices}
\end{equation}
This may be easily proven in the Hamiltonian representation of the lattice, by decomposing the $m_1$ summation as $m_1\to 2m_1'+H_1'$ in terms of a new summation variable $m_1'\in\mathbb Z$ and a new $\mathbb Z_2$ parameter $H_1'$. Subsequently, we introduce a Lagrange multiplier $G_1'$ (coupled to $n_1' \in\mathbb Z$) as a projector that allows the replacement $n_1\to n_1'/2$, and finally shift $n_1'\to n_1' -X_1-\gamma_2$ to remove the $X_1+\gamma_2$ terms from the lattice momenta. The resulting expression is a lattice with winding shift along the first direction (and momentum shifted along the second) at $T'=i\alpha'$ and $U'=2i$, which may be T-dualised back into our standard momentum-shifted lattice along both directions at $T=2i\alpha'$ and $U=i$. Making use of \eqref{IdentityLattices} (and dropping the primes) sends the summation parameters $\gamma_2,\delta_2$ of \eqref{Case3mix2} into the phase, and we can sum them to obtain
\begin{equation}
	\frac{1}{2^2}\sum_{H_1, H_2\atop G_1, G_2} (-1)^{(G_1 X_2+H_1 Y_2+G_1 H_1)+(G_2 X_1+H_2 Y_1+G_2 H_2)+(G_1 H_2+G_2 H_1)+(Y_1 X_2+X_1 Y_2)+K} \,\Gamma_{2,2}\left[\begin{array}{cc}H_1 & H_2  \\ G_1 & G_2 \end{array}\right](2i\alpha', i)\,.
	\label{Case3mix3}
\end{equation}
This is almost identical to the r.h.s. of \eqref{Case3mix1}, except that the phase exponent now contains the additional term $K=(H_2+X_2)(G_2+Y_2)$. It should be stressed that the r.h.s. of \eqref{Case3mix1} and \eqref{Case3mix3} are equal (as they should), despite the difference in the phase, but this equality is non-trivial. It is not a ``numerical" coincidence, but can be actually traced to the structure of the lattices.

Looking at these contributions in the context of the full partition function illustrates how different phases may still yield equivalent representations of the same theory. This is because the summation over the parameters $H_I,G_I$, associated to the freely-acting orbifold shifts, ultimately modify the choice of lattice momenta and the equivalent phases reflect the natural redundancy in picking the fundamental cell. Indeed, starting from either representation \eqref{Case3mix1} or \eqref{Case3mix3}, corresponding to the aforementioned equivalent choices of phase, and summing over $H_{1,2}$ and $G_{1,2}$ yields identical lattices, although some Poisson resummations (and T-dualities) may be needed in order to highlight this fact.

In the remainder of this section we shall explicitly work out this equivalence at the level of the lattices. Starting from the r.h.s. of \eqref{Case3mix1}, observe that the point $T=2i\alpha'$, $U=i$ corresponds to a square 2-torus without $B$-field, and with both radii sitting at the dual fermionic point, $R_1=R_2=\sqrt{2\alpha'}$. The (2,2) momentum-shifted lattice can then be cast into its Lagrangian representation (involving only windings) by starting from the Hamiltonian representation \eqref{ShiftedLatticeDef} and Poisson resumming over the momenta $m_i$,
\begin{equation}
	\Gamma_{2,2}\left[\begin{array}{cc}H_1 & H_2  \\ G_1 & G_2 \end{array}\right](2i\alpha', i) = \frac{(\sqrt{2\alpha'})^2}{\alpha'\tau_2} \sum_{m_i,n_i \in \mathbb Z} e^{ -\frac{\pi(\sqrt{2\alpha'})^2}{\alpha'\tau_2} \left[ \left|m_1+\frac{G_1}{2}+\tau\left(n_1+\frac{H_1}{2}\right)\right|^2 + \left| m_2+\frac{G_2}{2}+\tau\left(n_2+\frac{H_2}{2}\right)\right|^2 \right] } \,.
\end{equation}
In this form\footnote{Note that we still denote the summation variables by $m_i$ also in the Lagrangian representation, although these are now windings.}, the modularity properties of the shifted lattice are manifest at the expense of obscuring T-duality. The latter can be seen by a double Poisson resummation\footnote{The first Poisson resummation in the $m$'s brings the lattice back to its Hamiltonian form. A subsequent Poisson resummation in the $n$'s then brings it to its dual Lagrangian representation. Note that there is an exchange of notation $m_i\leftrightarrow n_i$ involved in this double Poisson resummation.}, 
\begin{equation}
	\Gamma_{2,2}\left[\begin{array}{cc}H_1 & H_2  \\ G_1 & G_2 \end{array}\right](2i\alpha', i) = \frac{(\sqrt{\alpha'/2})^2}{\alpha'\tau_2} \sum_{m_i,n_i \in \mathbb Z} e^{ -\frac{\pi(\sqrt{\alpha'/2})^2}{\alpha'\tau_2} \left[ \left|m_1+\tau n_1\right|^2 + \left| m_2+\tau n_2\right|^2 \right] } \,(-1)^{\textbf{m}\cdot\textbf{H}+\textbf{n}\cdot\textbf{G}} \,,
	\label{DualLagrangian}
\end{equation}
which has the advantage that it brings the freely-acting orbifold parameters $H_{1,2}$ and $G_{1,2}$ entirely into the phase. Plugging this into \eqref{Case3mix1} and summing over $H_1,G_1$ yields
\begin{equation}
	\begin{split}
	\frac{(\sqrt{\alpha'/2})^2}{\alpha'\tau_2} \sum_{m_i,n_i \in \mathbb Z} e^{ -\frac{\pi(\sqrt{\alpha'/2})^2}{\alpha'\tau_2} \left[ \left|m_1+\tau n_1\right|^2 + \left| m_2+\tau n_2\right|^2 \right] } \,(-1)^{m_1 X_2+n_1 Y_2} \,(-1)^{X_1 Y_2+Y_1 X_2+X_2 Y_2} \\
		\times \frac{1}{2}\sum_{H_2,G_2} (-1)^{H_2(m_1+m_2+Y_1+Y_2)+G_2(n_1+n_2+X_1+X_2)}  \,.
	\end{split}
	\label{Case3mix4}
\end{equation}
The summation over $H_2, G_2$ produces two projectors that impose $m_1+m_2+Y_1+Y_2=n_1+n_2+X_1+X_2=0$ (mod 2). We can solve both of these constraints in terms of a pair of new (unconstrained) windings $m_1',n_1' \in \mathbb Z$, by writing
\begin{equation}
	m_1 = 2m_1' + m_2+ Y_1+ Y_2 \ , \quad n_1 = 2n_1'+n_2+X_1+X_2 \,.
\end{equation}
Plugging this into \eqref{Case3mix4} and double Poisson resumming (first over $m_1'$ and then over $n_1'$ and dropping the primes) finally gives
\begin{equation}
	\begin{split}
	\frac{(\sqrt{\alpha'/2})^2}{\alpha'\tau_2} \sum_{m_i,n_i \in \mathbb Z} e^{ -\frac{\pi(\sqrt{\alpha'/2})^2}{\alpha'\tau_2} \left[ \left|m_1+\tau n_1\right|^2 + \left| m_2+\tau n_2\right|^2 \right] } \,(-1)^{m_1 (X_1+X_2)+n_1 (Y_1+Y_2)+m_2 X_1+n_2 Y_1+m_2 n_2}\, \\
		\times (-1)^{m_1 n_2+m_2 n_1} \,(-1)^{X_1 Y_2+Y_1 X_2+X_2 Y_2}  \,.
	\end{split}
	\label{Case3mix5}
\end{equation}
This is the expression we were looking for: the orbifold summation parameters $H_I,G_I$ are no longer present, and we can read off the coupling of the lattice windings to $X_{1,2}, Y_{1,2}$ from the phase. The contribution $(-1)^{m_1 n_2+m_2 n_1}$ corresponds to a $B$-field on the (square) 2-torus, at the special value $B_{12}=1/2$, while the radii are both at the fermionic point $\sqrt{\alpha'/2}$. Now consider instead the representation \eqref{Case3mix3} with the modified phase as the starting point. Working in a similar fashion, expressing the lattice in its dual Lagrangian representation \eqref{DualLagrangian}, and then performing the summation first over $H_1,G_1$ and then over $H_2, G_2$, directly leads to the very same lattice expression \eqref{Case3mix5}.
This proves the equivalence of the two phase representations we encountered in our simple toy model setup, but the main lesson is general. In more realistic models one typically has to deal with several basis vectors and independent shift directions, similar to  those encountered in Section \ref{SymExample}. In those cases, the redundancy in picking equivalent phases is naturally much larger, but the corresponding lattices match when brought to the form \eqref{Case3mix5}.

Notice furthermore, that shifting the lattice sum \eqref{Case3mix5} by $m_2\to m_2+m_1$ and $n_2\to n_2+n_1$, introduces a non-diagonal lattice metric and removes the $B$-field (while modifying the coupling to $X_{1,2}, Y_{1,2}$). If we then change the summation variables as $m_I = 2m_I' +G_I'$ and $n_I = 2n_I'+H_I'$, one recovers the shifted (2,2) lattice at the point $(T,U)=(2i\alpha', \frac{1+i}{2})$, but the ``discrete torsion" phase $(-1)^{H_1' G_2'+G_1' H_2'}$ is now absent. In more general cases where we also allow for non-vanishing Wilson lines, the redundancy will again be enlarged (\emph{c.f.} \cite{Ploger:2007iq} and a relevant discussion in \cite{Athanasopoulos:2016aws}).

	
\section*{Acknowledgments}	

We would like to thank Carlo Angelantonj for valuable discussions. Furthermore, we are grateful to the Centre de Physique Th\'eorique de l'Ecole Polytechnique  and the Laboratoire de Physique de l'Ecole Normale Sup\'erieure for their hospitality during various stages of this work.


\bibliographystyle{utphys}

\providecommand{\href}[2]{#2}\begingroup\raggedright\endgroup

\end{document}